\documentclass{aa}
\usepackage{epsfig,graphicx,lscape,subfigure}
\usepackage{float,longtable,comment}
\usepackage[none,bottom,light,timestamp]{draftcopy}
\usepackage{rotating}
\usepackage{natbib}
\newcommand{\kms}{{km\,s}$^{-1}$}

\newcommand{\Msun}{\,$\rm{M}_\odot$}
\newcommand{\vsini}{$v_{\rm{e}} \sin i$}
\newcommand{\ve}{$v_{\rm{e}}$}

\begin{document}

\title{The VLT-FLAMES Tarantula Survey X: Evidence for a bimodal distribution of rotational velocities for the single early B-type stars}

\author{P.L. Dufton\inst{1}, N. Langer\inst{2}, P.R. Dunstall\inst{1}, C.J. Evans\inst{3}, I. Brott\inst{4}, S.E. de Mink\inst{5,6}\thanks{Hubble Fellow}, I.D. Howarth\inst{7}, M. Kennedy\inst{1},   C. McEvoy\inst{1},  A.T. Potter\inst{8}, O.H. Ram\'irez-Agudelo\inst{9}, H. Sana\inst{9}, S. Sim\'on-D\'iaz\inst{10,11}, W. Taylor\inst{12}, J.S. Vink\inst{13}}
{\institute{Astrophysics Research Centre, School of Mathematics and Physics, Queen's University Belfast, Belfast BT7 1NN, UK
	 \and{Argelander Institut f\"{u}r Astronomie der
Universit\"{a}t Bonn, Auf dem H\"{u}gel 71, 53121 Bonn, Germany}
	\and {UK Astronomy Technology Centre, Royal Observatory Edinburgh, Blackford Hill, Edinburgh, EH9 3HJ, UK}
	\and{University of Vienna, Department of Astronomy, T\"{u}rkenschanzstr. 17, A-1180 Vienna, Austria}
        \and {Space Telescope Science Institute, 3700 San Martin Drive, Baltimore, MD 21218, USA}
        \and{Department of Physics and Astronomy, Johns Hopkins University, 3400 North Charles Street, Baltimore, MD 21218, USA}
        \and{Department of Physics and Astronomy, University College London, Gower Street, London, WC1E 6BT, UK}
        \and{Institute of Astronomy, The Observatories, Madingley Road, Cambridge CB3 0HA}
	\and{Astronomical Institute `Anton Pannekoek', University of Amsterdam, Postbus 94249, 1090 GE, Amsterdam, The Netherlands}
	\and{Instituto de Astrof\'isica de Canarias, E-38200 La Laguna, Tenerife, Spain}              
        \and{Departamento de Astrof\'isica, Universidad de La Laguna, E-38205 La Laguna, Tenerife, Spain}
	\and {Scottish Universities Physics Alliance, Institute for Astronomy, University of Edinburgh, Royal Observatory Edinburgh, Blackford Hill,
                 Edinburgh, EH9 3HJ, UK}
       \and{Armagh Observatory, College Hill, Armagh BT61 9DG, Northern Ireland, UK}}
}
       

\date{Received; accepted }

\abstract{}
{Projected rotational velocities (\vsini) have been estimated for 334 targets in the VLT-FLAMES Tarantula survey that do not manifest significant radial velocity variations and are not supergiants. They have spectral types from approximately O9.5 to B3. The estimates have been analysed to infer the underlying rotational velocity distribution, which is critical for understanding the evolution of massive stars.} 
{Projected rotational velocities were deduced from the Fourier transforms of spectral lines, with upper limits also being obtained from profile fitting. For the narrower lined stars, metal and non-diffuse helium lines were adopted, and for the broader lined stars, both non-diffuse and diffuse helium lines; the estimates obtained using the different sets of lines are in good agreement. The uncertainty in the mean estimates is typically 4\% for most targets. The iterative deconvolution procedure of Lucy has been used to deduce the probability density distribution of the rotational velocities.}
{Projected rotational velocities range up to approximately 450 \kms and show a bi-modal structure. This is also present in the inferred rotational velocity distribution with  25\% of the sample  having $0\leq$\ve$\leq$100\,\kms and the high velocity component having \ve$\sim 250$\,\kms.  There is no evidence from the spatial  and radial velocity distributions of the two components that they represent either field and cluster populations or different episodes of star formation. Be-type stars have also been identified.
}
{The bi-modal rotational velocity distribution in our sample resembles that found for late-B 
and early-A type stars. While magnetic braking appears to be a possible mechanism for producing the low-velocity component, we can not rule out alternative explanations.}


\authorrunning{P.L. Dufton et al}
\titlerunning{Rotational velocities of B-type stars in the Tarantula Nebula}

\maketitle
%
\section{Introduction}                                         \label{s_intro}

Rotation is thought to have  a major influence on the evolution of massive stars. It may affect both how the star evolves during hydrogen core burning \citep[see, for example,][]{bro11a, pot12a} and  the nature of any supernova explosion \citep[see, for example][]{woo06, yoo06}. It can also provide insights into other phenomena such as star formation \citep[see, for example,][]{wol07} and how magnetic fields affect stellar evolution \citep{pot12b}. 

There have been several large studies of rotation in massive main sequence stars both in our own Galaxy and in the Magellanic Clouds. For the former, one of the earliest comprehensive samples was provided by \citet{sle49}. As this sample was selected without any previous knowledge about the stellar spectra, it provides an unbiased sample. Estimates of the projected rotational velocities of 123 early-type stars were presented together with upper limits for 63 stars with relatively narrow absorption lines. \citet{abt02} investigated 1100 B-type stars selected from the bright star catalogue and this is one of the largest samples for which the biases are well understood. For late B-type  stars they find evidence for a bi-modal distribution but early-B type stars do not seem to show this bi-modality, although the number of such stars in their sample is small. \citet{hua10}  combined data from their observing campaigns to provide projected rotational velocities for approximately 660 cluster stars and 470 field stars. They find that the average projected rotational velocity is significantly lower for field stars, which they attribute to the fact that they are systematically more evolved and spun down than their cluster counterparts. 

For the Magellanic Clouds, there have been two investigations using ESO large programmes, viz. \citet[][and references therein]{mar06, mart07} and \citet[][hereafter designated FLAMES-I]{hun08a, eva05, eva06}. These observed predominantly field stars, while as discussed above for the Galactic studies, the behaviour of cluster stars may be different. Also given the limited time cadence of the observations, it was not always possible to reliably distinguish between single and multiple systems. 

Here we present projected rotational velocity estimates for a sample of 334 early B- and late O-type stars located in the 30 Doradus region \citep[][hereafter Paper I]{eva11}. The rotational velocities of the more massive O-type stars will be discussed elsewhere (Ramirez-Agudelo et al. in preparation). Our sample has been selected as having no significant radial velocity variations, thereby reducing the contamination from multiple systems. We have also inferred the intrinsic rotational velocity distribution and discuss its bi-modal structure. 

\section{Observations}                                        \label{s_obs}

The spectroscopic data were obtained as part of the European Southern Observatory (ESO) large programme \citep{eva11} using the Fibre Large Array Multi-Element Spectrograph, FLAMES \citep{pas02} on the Very Large Telescope (VLT), primarily with the Giraffe spectrograph, but also supplemented with data from the Ultraviolet and Visual Echelle Spectrograph, UVES \citep{dek00}.  All observations have been based in and around the LMC open association 30 Doradus, with the core region R136 having a 15 arcsecond exclusion radius as it is too densely populated for use of the Medusa fibres.  This region has been observed with five pointings using the ARGUS integral field unit (IFU).  The current work will use the observations obtained using the two bluer region Giraffe spectra to obtain estimates of the projected rotational velocity and the red spectra to search for Be-type emission. The observational configurations are summarized in Table \ref{t_obs}, with details of the target selection,  observations and data reduction being available in Paper I. 

The sample is that defined in Paper I as potentially B-type stars and comprises approximately 540 targets.  Ongoing spectral classification indicates that some targets are late-O type stars with spectral types of O9.5 or O9.7. It is believed that our sample is effectively complete as regards the O9.7 and later spectral types but is incomplete for the O9.5 spectral type as approximately half these targets were included only in the dataset of potentially O-type stars. We return to this in Sect. \ref{d_gen}. Additionally the magnitude cutoff employed in the survey implies that subject to spatial variations in extinction our sample should not include (near) zero-age  main sequence stars with spectral types later than approximately B3.

\citet{dun11} have undertaken a radial velocity analysis for this dataset to identify binaries. They find 37\% of their sample show unambiguous velocity variations with an additional 6\% having an uncertain binary status. We excluded the former from our analysis but included the latter, although we consider these results separately. Additionally thirty stars without significant radial velocity variations have been identified as supergiants and a preliminary model atmosphere analysis, similar to that employed by \citet{fra10} implied that they all had logarithmic gravities between 2.5 and 3.2 dex. As these objects will normally have evolved from O-type stars, they will be discussed elsewhere (McEvoy et al, in preparation; Sim\'on-D\'iaz et al., in preparation).

Hence our sample of 334 stars (of which 45  are of uncertain status) is a subset of the Tarantula B-type sample that lie close to the main sequence. They show no convincing evidence of any radial velocity variation typically at the 10 \kms\  level over six epochs spread over one year. Assuming a spectral type range of B3 to O9.5, leads to an effective temperature range of approximately 20\,000 to 34\ 000\,K \citep{tru07} and a mass range of approximately 6 to 16 \Msun \citep{bro11a} for an LMC metallicity. 

\citet{dun11} have discussed completeness and estimate that up to 50\% of the apparent B-type main sequence sample could be undetected binaries. Indeed the failure to identify long period systems (typically $P>1000$~days) is an intrinsic limitation of the VLT-FLAMES Tarantula Survey (VFTS) given the one year baseline of the observations. Assuming that this analysis is correct, our sample could contain a significant fraction of binaries, albeit preferentially with long orbital period and/or low-mass companions. The light and spectrum from such binaries would be dominated by the main (B-type) component. Additionally the long orbital periods involved are unlikely to have led to any significant tidal interaction that would have modified the rotation of the objects. 

As there were no significant radial velocity variations in our sample, spectra could be combined without including any wavelength shifts. Several methods were adopted with the earlier analysis being based on co-added spectra with exposures containing cosmic rays or having a low signal-to-noise (SNR) being excluded. Later, all the exposures were normalised using low order polynomials. Normally an extended wavelength region could be normalised with a single polynomial but for some exposures, there was significant variations in the continuum flux with wavelength (se Paper I for details). In these cases individual lines or groups of lines were normalised. Spectra were then combined with a simple median filter or a weighted mean with $\sigma$-clipping; in both cases low SNR exposures were again excluded. Given the relatively small range in SNR for a given star and wavelength region, these two approaches gave essentially identical spectra. Comparison of these data and of the earlier co-added spectra for five stars showed excellent agreement (at the 1-2\% level) in the estimated mean projected rotational velocities indicating that the choice of reduction technique should not be a significant source of uncertainty.

\begin{table}
\caption{Summary of the spectroscopy used in the rotational velocity analysis. For some stars additional LR02 observations were obtained when the initial observations were found to be of poor quality due to weather conditions.}
\label{t_obs}
\begin{center}
\begin{tabular}{lrrr}
\hline\hline
Instrument		&	$\lambda$-range &	R 		&	Exposures 
\\
                                &     (\AA)                       &                 &      (seconds)
\\
\hline
Giraffe LR02		&	3960-4560	&	7000 	&	6 x (2 x 1815)\\
Giraffe LR03		&	4505-5050	&	8500	&	3 x (2 x 1815) \\
Giraffe HR15N		&	6445-6815	&	16000	&	2 x (2 x 2265) \\
\hline
\end{tabular}
\end{center}

\end{table}

\section{Projected rotational velocities} 	

\subsection{Methodology}\label{s_ft}
Several methods are available for estimating projected rotational velocities (\vsini). Those most commonly used for OB-type stars include a direct measure of the full-width-half-maximum of the spectral lines \citep[see, for example][]{sle75,abt02,str05}, profile fitting \citep[see, for example,][]{rya02,hun08a} or cross correlation  \citep[see, for example,][]{how97}. \citet{sim07} have discussed the utility of the Fourier transform (FT) method for early type stars, as this should be able to separate the rotational broadening from other broadening mechanisms such as macroturbulence. Recently this approach has been widely used to study the different mechanisms contributing to the broadening of spectral lines in early-type supergiants \citep{dufsmc06, lef07, mar07, sim10, fra10}. 

\begin{figure}
   \centering
   \includegraphics[scale = 0.24,angle=0]{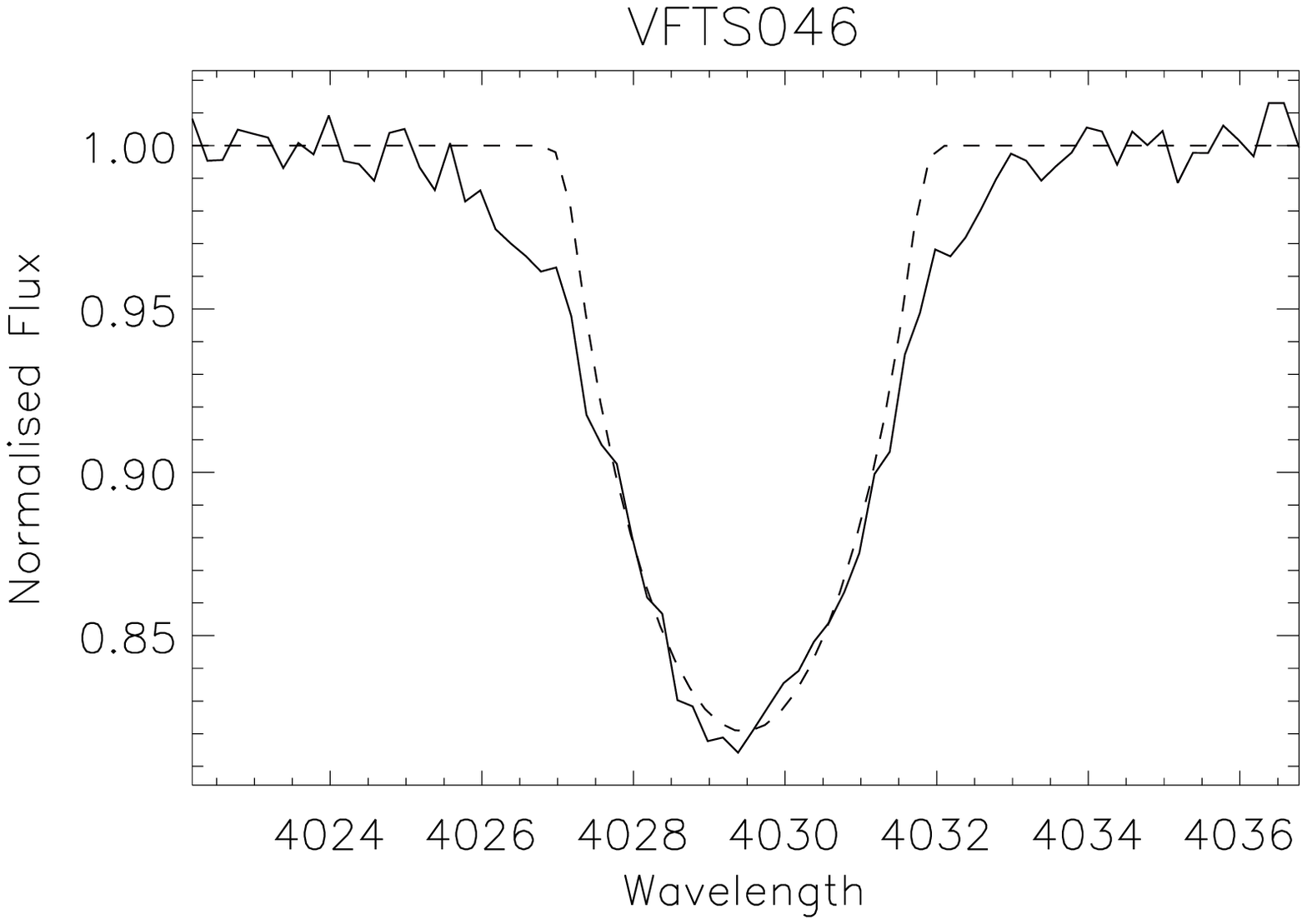}
   \includegraphics[scale = 0.24,angle=0]{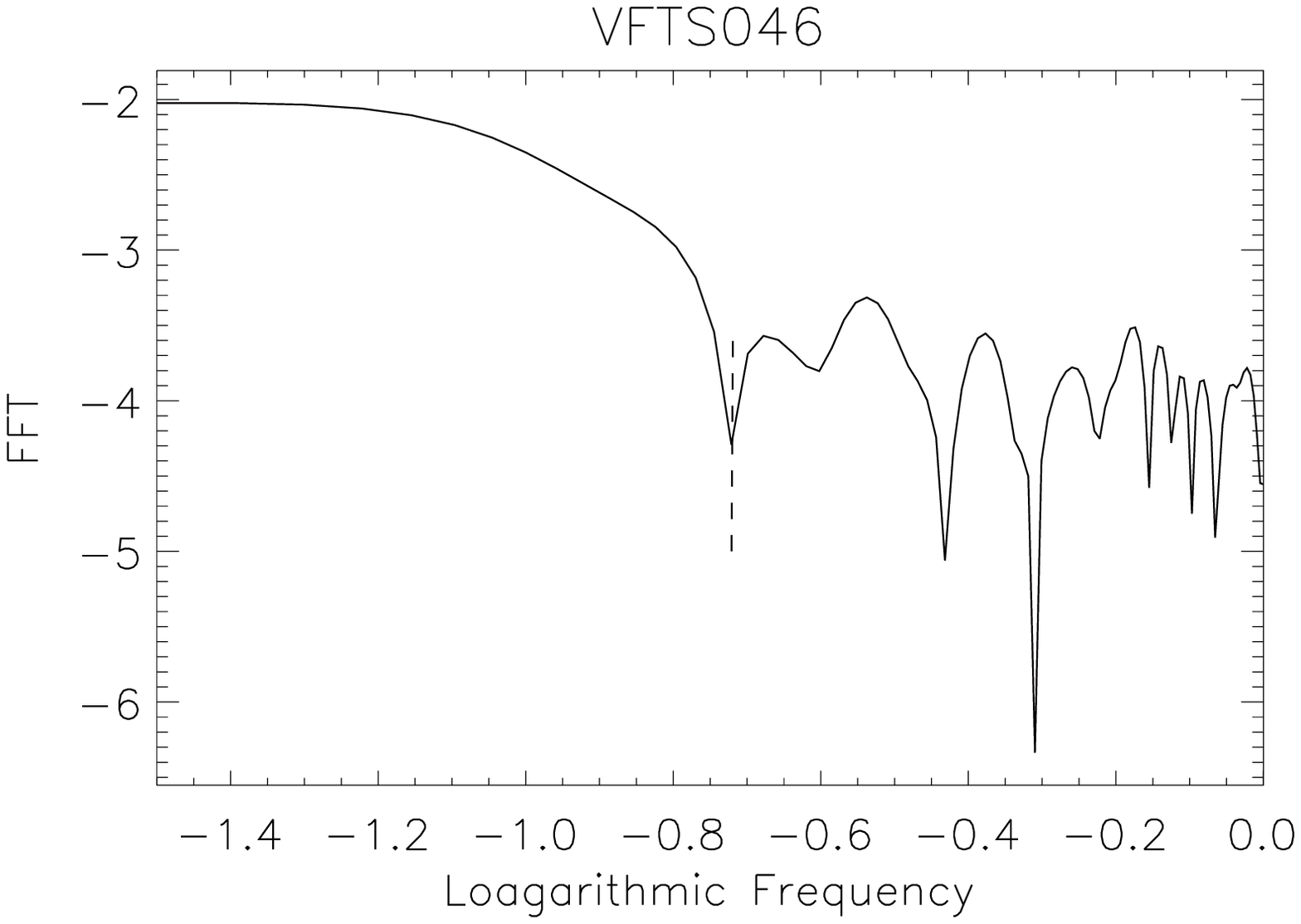}\\
   \includegraphics[scale = 0.24,angle=0]{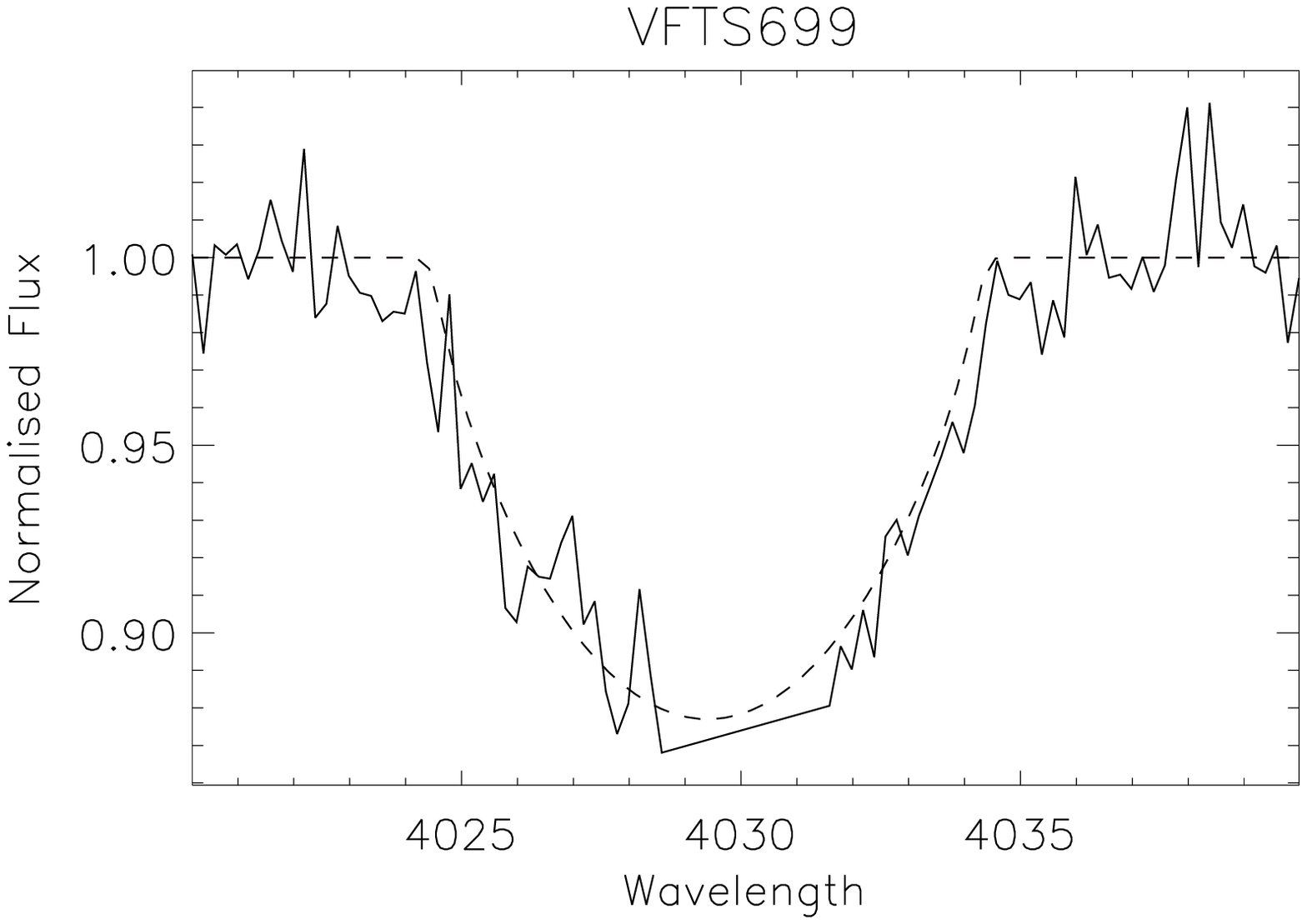}
   \includegraphics[scale = 0.24,angle=0]{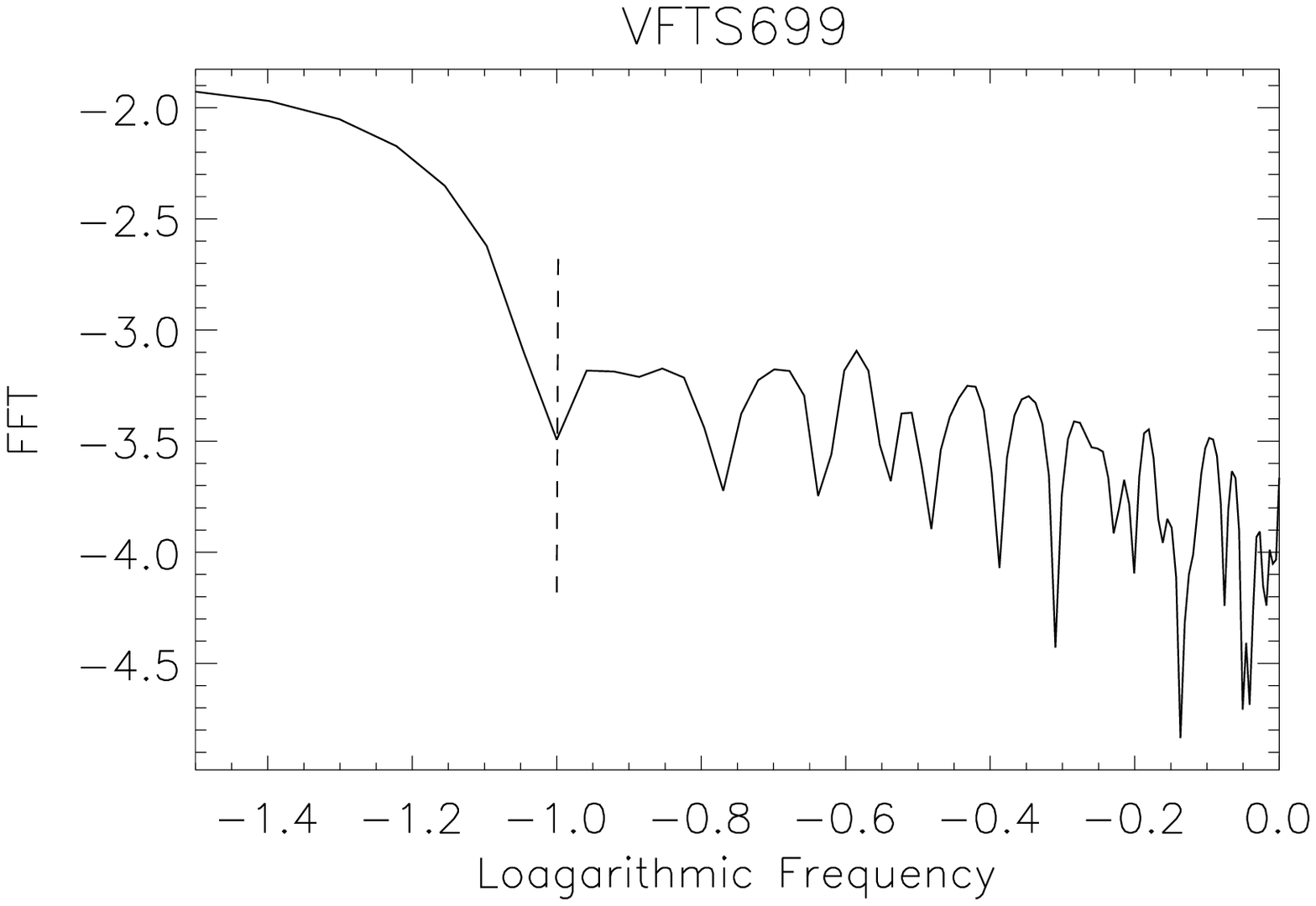}
   \caption{Profiles and Fourier transforms for the \ion{He}{i} line at 4026\AA\ in VFTS046 and VFTS699. The theoretical profiles correspond to the \vsini\ estimates from the Fourier Transform methodology. The position of the first mimium in the Fourier Transform is marked by a vertical line. For VFTS699, it has been necessary to remove some nebular emission near the line centre  by linear interpolation over the affected wavelength region.}
   \label{FT_ex}
\end{figure}

We have adopted this Fourier transform  method for all the stars in our sample. This approach relies on identifying the first minimum in the Fourier transform for a spectral line, which is assumed to be the first zero in the Fourier transform of the rotational broadening profile with the other broadening mechanisms exhibiting either no minima or only minima at higher frequencies. Fig. \ref{FT_ex} shows the \ion{He}{i} line at 4026\AA\ and its Fourier Transform for two targets, VFTS046 and VFTS699. Their  \ion{He}{ii} spectra are of similar strength (implying a spectral type near B0) and hence the difference in their line profile reflects their different projected rotational velocities. Also shown are the  rotational broadening profiles for a projected rotational velocity, corresponding to the first minimum in the Fourier Transform. As would be expected these agree well with the observations in the line centre but are narrower in the wings for VFTS046 due to the intrinsic Stark broadening of this diffuse line.

As discussed by \citet{sim07}, the first minimum in the Fourier Transform can be difficult to identify in spectra with, for example, low SNRs or significant nebular contamination.  Therefore we have also profile fitted (PF) all the spectral features,  assuming that rotation dominates the line broadening i.e. we have used a simple rotational profile and assumed that the intrinsic and instrumental profiles can be approximated by scaled $\delta$-functions.  The neglect of other broadening mechanisms will lead to these estimates being upper limits but they provide a useful constraint on the actual projected rotational velocities. We discuss these results further in \ref{s_comp}.

Two different sets of spectral lines were adopted for analysis depending on the degree of rotational broadening. For the narrower lined stars (with \vsini\  $\la 150$\kms), a selection of metal lines and non-diffuse helium lines (Set 1 in Table \ref{t_lines}) was used. These had the advantage of being intrinsically narrow, whilst the former were not normally affected by nebula emission. For large rotational velocities, reliable results could only be obtained from stronger absorption lines and in these cases a selection of diffuse and non-diffuse helium lines (Set 2 in Table \ref{t_lines}) was considered. Some of the features are in fact close doublets or triplets, whilst additionally the helium diffuse lines have weak forbidden components (we note that blending of the \ion{He}{i} 4026\AA\ feature with a \ion{He}{ii} line should not be significant for our sample).  For the former, the separations were less than the broadening due to rotation and indeed the instrumental profile, whilst the latter were only used for the more rapidly rotating stars where the profiles appear symmetric within the observational uncertainties.

Normally it was not possible to obtain estimates from all the spectral lines in the adopted set for three main reasons. Firstly, some lines were too weak to provide reliable measurements. For example, the \ion{C}{ii} and \ion{Mg}{ii} features could not be used for the hotter stars, whilst the \ion{O}{ii} and \ion{Si}{iii} features were weak in the cooler objects. Secondly, blending could affect some lines and in particular the \ion{He}{i} line at 4121\AA\ (blended with \ion{O}{ii}) and the \ion{Mg}{ii} line at 4481\AA\ (blended with  \ion{Al}{iii}). However the most serious constraint was the presence of strong nebular emission in the spectrum of some stars, which could not be reliably removed for our multi-fibre observations. This emission could affect the \ion{C}{ii} feature but more seriously the \ion{He}{i} spectrum. The following approach was taken for increasing amounts of contamination.

\begin{enumerate}
\item The emission was snipped out of the spectra (see Fig. \ref{FT_ex}), which was then regridded and analysed. The quality of the Fourier transform, the consistency between the estimates from the FT and PF approaches and the quality of the fit of the rotationally broadened profile to the observations was then used to assess the reliability of the estimate.
\item The lines with the greatest nebular contamination (normally the diffuse triplet transitions) were excluded. As can be seen from Table 4, there are no systematic differences between the estimates obtained from different lines. Hence the use of a subset of lines is unlikely to introduce significant biases.
\item When a reliable estimate could not be obtained, an attempt was made to assign the star to a velocity range covering  40 \kms\  (0-40, 40-80, 80-120 \kms\  etc.). The observed profiles were then compared with rotationally broadened profiles appropriate to this range to validate this decision. Hence although no reliable projected rotational velocity could be estimated, it would be possible to include such stars in any statistical study.
\end{enumerate}

There were a number of narrow lined stars whose line width appeared to be dominated by the instrumental profile which corresponded to a FWHM of approximately 40 \kms. As would be expected, their Fourier Transforms did not show convincing minima with individual measurements showing a significant degree of scatter. Other stars showed a greater degree of broadening and yielded mean projected rotational velocities in the range of 30 to 40 \kms. Here we have taken a conservative approach and have estimated rotational velocities only for stars with individual measurements from the Fourier Transform methodology that were consistently greater than 40 \kms\  and where the profile fitting implied a significant rotational broadening component. 

For each star, a projected rotational velocity was estimated only if there was at least three independent measurements. For the narrow lined stars, the average number of measurements was 6.9, with approximately 90\% having five or more measurements, with the corresponding values for the broader lined stars  being 6.3 and 93\%. For the latter, an additional 16 stars were assigned projected rotational velocity ranges whilst  one star, VFTS729, appeared to have asymmetric line profiles, suggesting a binary or composite nature, and was excluded. 

Tables 3  and 4 (only available online) lists all the individual measurements using both the FT and the PF methodologies, together with the percentage depth of the feature as estimated from the profile fitting. The former  is for the metal line set and the latter for the helium line set. In both tables stars that are of an uncertain status i.e. could possibly be binaries (see Sect. \ref{s_obs}) are identified with an asterisk. The tables also give means and standard deviations with the FT estimates being weighted by the central depths. Using unweighted means would have lead to no systematic shift in the estimates with typical changes of 2\% for line set 1,  and a 1\% decrease for line set 2 with typical changes again being 2\%. Hence the choice of weighting is unlikely to be a significant source of error. We also list  the mean values from the profile fitting; in this case the individual estimates were not weighted as these are better considered as providing upper bounds. Stars with narrow lined spectra that were instrumentally dominated have been assigned a projected rotational velocity of $\la 40$\,\kms.

 \citet{tow04} and \citet{fre05} have investigated the effects of equatorial gravity darkening due to   angular rotation ($\Omega$) on observed line profiles. They find that for angular velocities near the critical limit ($\Omega_c$) the effects can be significant and lead to a systematic underestimation of the real projected rotational velocity. For example, \citet{tow04} considered a representative metal and helium diffuse line in stars with a spectral range B0-B3 (similar to that observed here) and for $\Omega$/$\Omega_c$=0.95 and $\sin i = 1$. They found the projected rotational velocity would be underestimated by between 9 and 22\%. Unfortunately we do not know either the $\sin i$\ values or atmospheric parameters of our targets and are hence unable to systematically correct for this effect. 

However we can use the results of \citet{mar06} to obtain an insight into the magnitude of any systematic effects. They used the methodology of \citet{fre05} to correct observed projected rotational velocities for 47 Be-type stars towards the LMC cluster, NGC\,2004. For $\Omega$/$\Omega_c=0.85$\ (the lowest value that they considered), the correction were 5.5\% for stars with \vsini $<200$\kms, 3.6\% ($200\leq$\vsini$<300$\kms), 3.3\% ($300\leq$\vsini$<400$\kms) and 3.0\% ($400\leq$\vsini$<500$\kms). The systematic errors in our estimates should be considerably smaller as it is unlikely that the majority of our targets are rotating with such high values of $\Omega$/$\Omega_c$. For example, taking the median rotational velocity estimated from the de-convolution discussed in Sect. \ref{d_decon} and the critical velocities from the grid of \citet{bro11a} for early B-type stars leads to a value of $\Omega$/$\Omega_c\simeq 0.3-0.4$. Hence we do not expect that there will be a significant effect on our sample as a whole, although a small number of individual estimates (where the target is rotating with near critical velocity) could be affected.

\begin{table}
\caption{Sets of absorption lines used for projected rotational velocity estimates}\label{t_lines}
\centering
\begin{tabular}{cccccccc}
\hline\hline
\multicolumn{2}{c}{Set 1} & \multicolumn{2}{c}{Set 2} &  \\
\hline
Species & $\lambda$\ (\AA) & Species & $\lambda$\ (\AA) \\
\ion{He}{i}   & 4120  & \ion{He}{i} & 4009 &  \\
\ion{He}{i}   & 4169  & \ion{He}{i} & 4026 &  \\
\ion{C}{ii}    & 4267  & \ion{He}{i} & 4120 &  \\
\ion{He}{i}   & 4437  & \ion{He}{i} & 4143 &  \\
\ion{Mg}{ii}  & 4481  & \ion{He}{i} & 4387 &  \\
\ion{Si}{iii}   & 4552  & \ion{He}{i} & 4471 &  \\
\ion{Si}{iii}   & 4567  & \ion{He}{i} & 4713 &  \\
\ion{Si}{iii}   & 4574  & \ion{He}{i} & 4921 &  \\
\ion{O}{ii}    & 4661  &      -           &  -       &  \\
\ion{He}{i}   & 4713  &      -           &  -       &  \\

\hline
\end{tabular}
\end{table}

\subsection{Results and estimated uncertainties} \label{s_comp}

Before discussing the significance of the results presented in Tables 3 and 4, it is important to validate their reliability. Several tests were undertaken as discussed below. Firstly one star (VFTS331) was inadvertently analysed twice with excellent agreement; the mean of the FT estimates agreed to within 2\,\kms\  and that of the PF estimates to within 3 \kms. Secondly four targets in the overlap region between the narrow and broad lines stars ($100\la$\vsini$\la 200$\,\kms) were measured using both sets of lines and  in Table  \ref{t_comp}  the mean values and the estimated errors are summarized.  The estimates are in reasonable agreement with no evidence of systematic differences.  The t-statistic and degrees of freedom for  unequal sample sizes and unequal variance (Welch's t-test) are also listed. None of these are significant at the 10\% levels, but those for VFTS234 and VFTS452 are significant at at the 20\% level.

\addtocounter{table}{+2}
\begin{table}
 \caption{Comparison of mean projected rotational velocities (\kms) deduced using the Fourier Transform methodology from the two line sets. Also listed are the standard deviations and number (in brackets) of the estimates. The modulus of the t-statistic and number of degrees of freedom are also listed.}
\label{t_comp}
\begin{center}
\begin{tabular}{lccccc}
\hline\hline
Star		&     Line set 1	          & Line set 2 & $|t|$ & df\\
\hline
046          &     169$\pm$15 (7)      & 171$\pm$15 (7) & 0.2 & 6\\
060          &     129$\pm$16 (9)     & 129$\pm$8    (7) & 0.0 & 6\\
234          &     157$\pm$12 (7)     & 147$\pm$6    (7) & 1.9 & 6\\
452          &     125$\pm$15 (7)     & 134$\pm$5    (8) & 1.5 & 7\\
\hline
\end{tabular}
\end{center}

\end{table}

\begin{figure}
   \centering
   \includegraphics[scale = 0.5,angle=0]{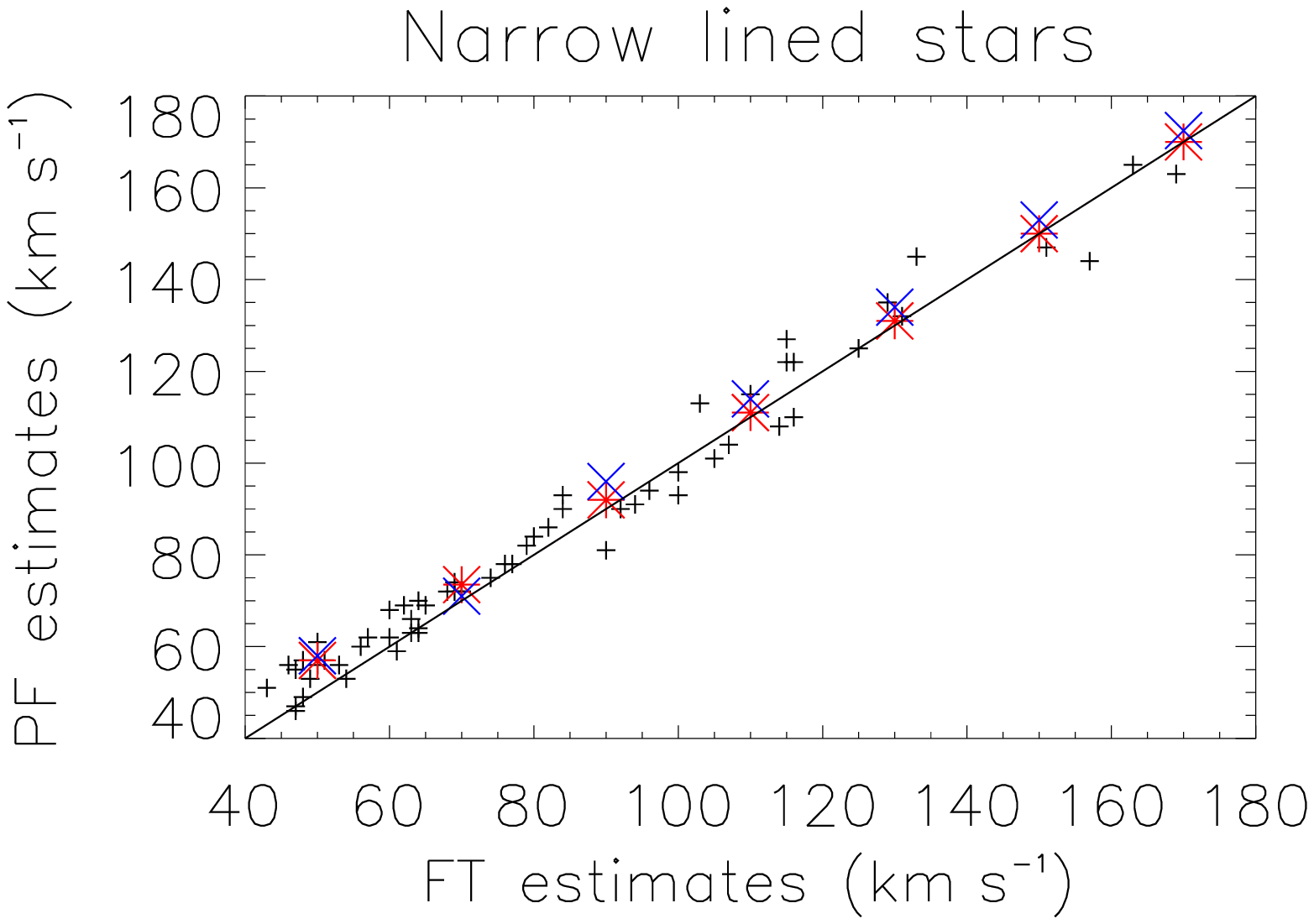}
   \includegraphics[scale = 0.5,angle=0]{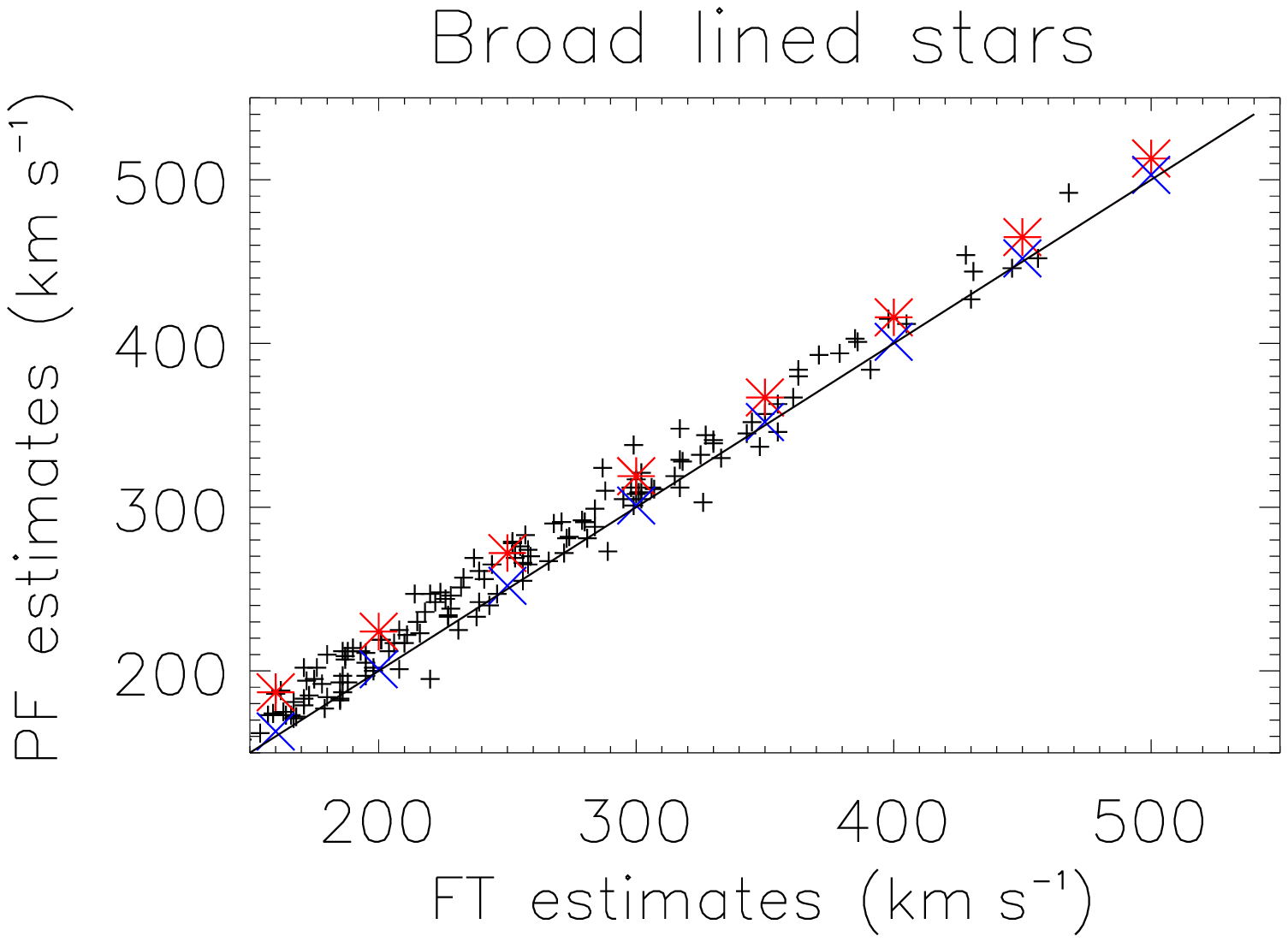}
   \caption{Comparison of projected rotational velocities estimated using the FT and PF methodologies. The upper figure is for the narrower lined stars (line set 1), whilst the lower figure is for the broader lined (line set 2) sample. Also shown are the results of the simulation -- crosses: \ion{He}{i} 4713\AA; asterisks: \ion{Si}{iii} 4552\AA\ or \ion{He}{i} 4387\AA.}
   \label{f_comp}
\end{figure}

A comparison has also been made of the mean projected rotational velocities obtained from the FT and PF methodologies. These are illustrated in Fig. \ref{f_comp}  for the the narrow lined (line set 1) and broader lined (line set 2) subsamples. In general the results are in excellent agreement for both line sets. At lower projected rotational velocities (e.g. $\la 80$\ \kms), the profile fitting methodology yields systematically higher estimates, which would be expected given that only rotational broadening was considered. 

To test this hypothesis, we took the spectrum from a TLUSTY model \citep{Hub88, Hub98}, with an effective temperature of 25\ 000 K, a logarithmic gravity of 4.0 dex and an LMC composition \citep[see][for details of the grid]{Rya03,Duf05}. This was convolved with a Gaussian profile to represent instrumental broadening and with different amounts of projected rotational broadening. The profiles were then analysed  using the PF method to estimate the projected rotational velocity. Three representative absorption features were considered, viz. the \ion{Si}{iii} line at 4552\AA, the non-diffuse helium line at 4713\AA\ and the diffuse helium line at 4387\AA. These results are also shown in  Fig. \ref{f_comp} and reproduce the trend in the differences between the observational estimates discussed above. Indeed although they only represent a subset of lines for a single model, they provide strong evidence for the reliability of our Fourier Transform results.

The sample standard deviations are typically 10\% of the mean projected rotational velocities for both line sets. However the error in the mean values will be smaller and if the estimates are normally distributed, this would be the sample standard deviation divided by the square root of the number of observations. This would then lead to a typical uncertainty in the mean values of 4\%. 

An independent error estimate can be found at higher projected rotational velocities, as here  the profile fitting methodology should provide reliable estimates. We have therefore compared the two methodologies in these regimes. For line set 1, we considered all stars that had estimated projected rotational velocities of greater than 80 \kms. For each star the ratio of the PF to the FT estimate was found, with the mean of these ratios being $1.01\pm 0.06$. For line set 2 and limiting our sample to those stars with estimates greater than 300 \kms\  leads to a mean ratio of  $1.02\pm 0.04$. Hence in these high projected rotational velocity regimes, the methodologies give excellent agreement, confirming the trends seen in Fig. \ref{f_comp}. Additionally as the standard deviations for the ratios will include a contribution from uncertainties in the PF approach, they are consistent with our estimated error of 4\% for the means of the FT estimates.

Rotational velocities have also been estimated for the apparently single O-type stars in the Tarantula Survey (Ramirez-Agudelo et al., in preparation). They also used the FT technique  but a different sets of lines reflecting the different temperature range. Due to uncertainties in the preliminary spectral classification, late O- and early B-type targets were included in both datasets (see Sect. \ref{s_obs}). Excluding targets that have only been assigned here to a velocity bin, there are 42 stars with independent estimates. There is no significant systematic difference between the two sets of measurements, with the mean value of the  absolute differences being 3$\pm$15 \kms and of the relative differences being 0$\pm$8\%. For thirty seven of the targets, the difference is less than 10\%, with the remaining five ranging from 10-30\%.  We discuss these discrepancy cases below:

{\bf VFTS 226, 627:} These targets have narrow lined spectra and are hence close to the resolution limit for getting reliable projected rotational velocities. Additionally macroturbulence may be significant given their relatively large effective temperatures. Hence we do not consider the relatively small absolute discrepancies (7-11\kms) in the estimates to be significant.

{\bf VFTS 373, 543:} Strong emission is present in both these spectra, making reliable estimates difficult. Indeed our sample standard deviations are approximately 30 \kms (see Table 4), implying that the estimates are consistent within the uncertainties. 

{\bf VFTS 141:} The spectrum of this star is well observed, with the nebular emission being weak. Although the two sets of measurements differ by 19\%, when only lines that are common to both analyses are included this decreases to 7\%. Hence we believe that the measurements are consistent.

\subsection{Distribution of projected radial velocities} \label{d_gen}

A histogram of our projected rotational velocities, excluding the targets with an uncertain status, is shown in Fig. \ref{f_hist}. A binsize of 40 \kms\  has been adopted as this is consistent with both our spectrally unresolved targets and those where only a velocity range was assigned.  The distribution appears to be double peaked with approximately 25\% of the sample, having \vsini$\leq 80$\kms. The sample should be relatively unbiased as it was selected on the basis of magnitude (with no colour cut-off) and then on the ability to  position the spectrograph fibres. The former could bias the sample towards targets with low extinction, whilst the latter could bias the sample away from regions with high star densities but neither of these would appear to be sufficient to explain the double peak.  Another explanation of the bi-modal structure could be that many of the narrow-lined `single'  stars are in fact part of binary systems with synchronised orbits. We would expect such systems to have relatively small orbital periods with large radial velocity variations, whilst their radial velocity estimates will have a high accuracy. As such the radial velocity analysis of \citet{dun11} should have identified such targets and hence this bias should not exist.

As discussed in Sect. \ref{s_obs}, we believe that our sample contains effectively all the targets with spectral types O9.7 or later but only partially samples those with spectral type O9.5. Hence in Fig. \ref{f_hist}, we also show a histogram with the O9.5 stars excluded. We note that the two histograms are very similar with any differences being less than the random Poisson noise expected from finite sample sizes. Hence in the subsequent discussion of these data, we have considered the complete sample but excluded stars with an uncertain binary status. Our principle conclusions would remain unchanged if we had excluded the targets with a O9.5 spectral type and/or included those of uncertain status. 

\subsection{Be-type stars}

 The classical Be-type star is a B-type star that shows or has previously shown prominent emission features in its Balmer line spectrum, indicating the presence of a geometrically flattened, circumstellar disc \citep{qui94, qui97}. Although the mechanism for the Be phenomena remains unclear, such stars are believed to rotate with velocities greater than 40\% \citep{cra05} up to approximately 90\% of the critical limit, where the centrifugal force balances that of gravity \citep{tow04}. 
Reliable identification of Be-type stars in this sample was difficult because of the strong nebular emission. Additionally sky subtraction did not work well for these fibre-fed spectra, because of the spatially highly variable nature of this emission. Hence to search for evidence of Be-type emission, we examined red-region (HR15N) spectra processed without sky subtraction.  Although there was no prospect of identifying weak or narrow stellar emission, broad Be-type H$\alpha$ emission wings were identifiable in 52 out of the 289 constant-velocity stars \citep[including VFTS 102, see ][]{duf11}, and 8 out of 45 stars of uncertain status -- i.e., about 18\%\ of the sample.  This fraction is comparable with canonical figures for LMC {\em clusters} (e.g., \citealt{mae99, wis06}), even though we are only able to identify relatively strong emission. The Be-type candidates are identified in Tables 3 and 4 by crosses (+).

\begin{figure}
   \centering
   \includegraphics[scale = 0.5,angle=0]{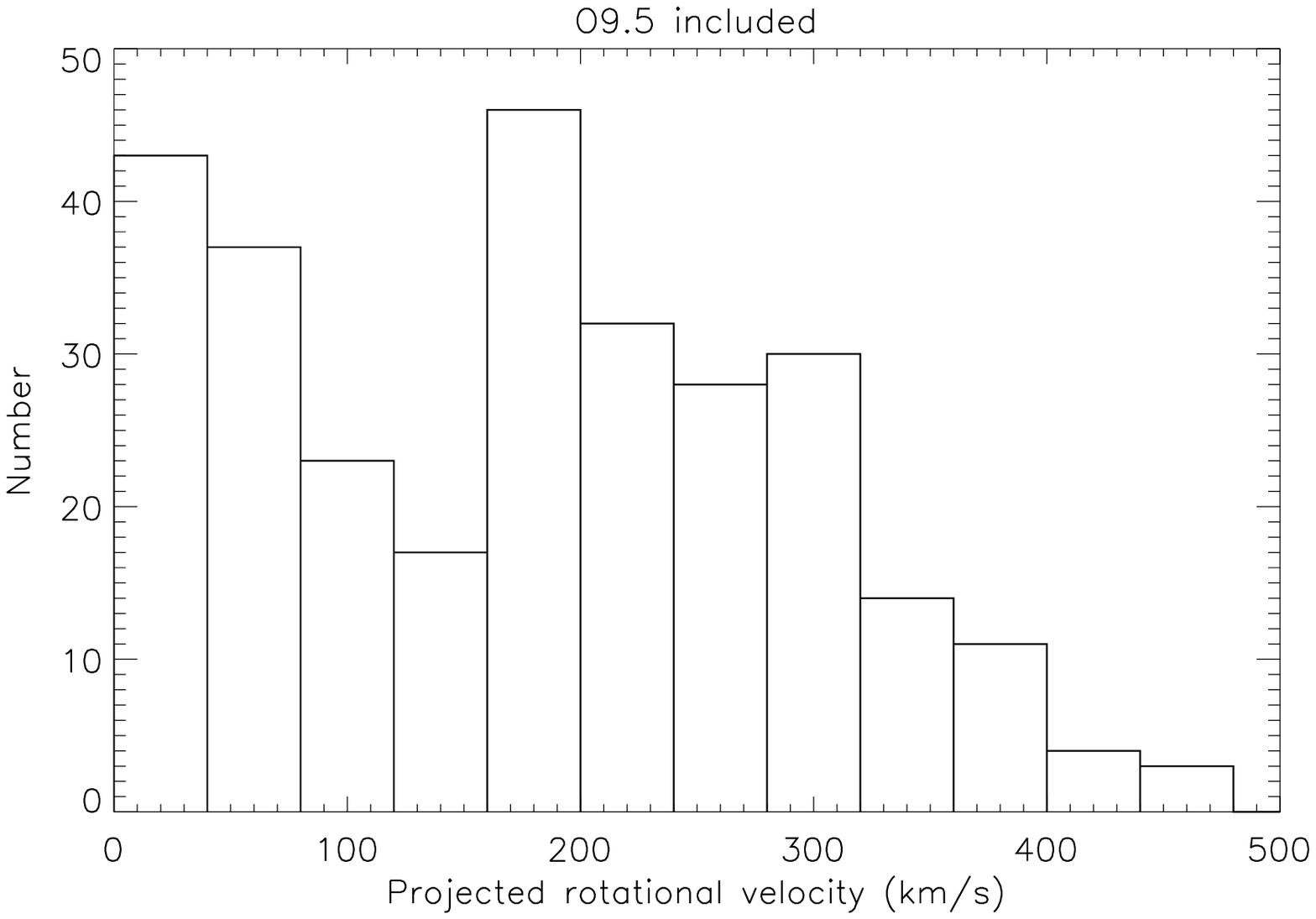}
   \\ 
   \includegraphics[scale = 0.5,angle=0]{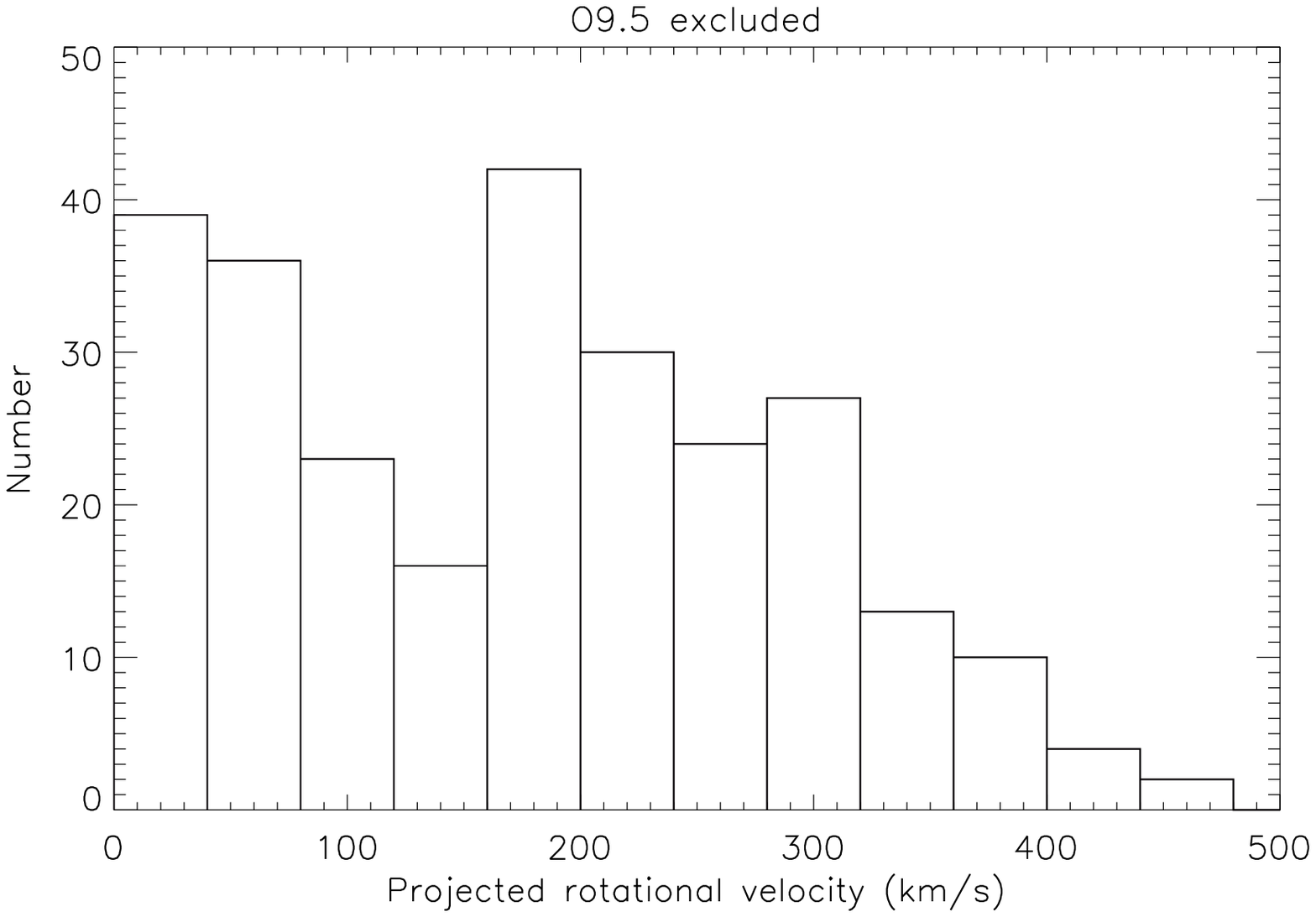}
   \\
   \includegraphics[scale = 0.5,angle=0]{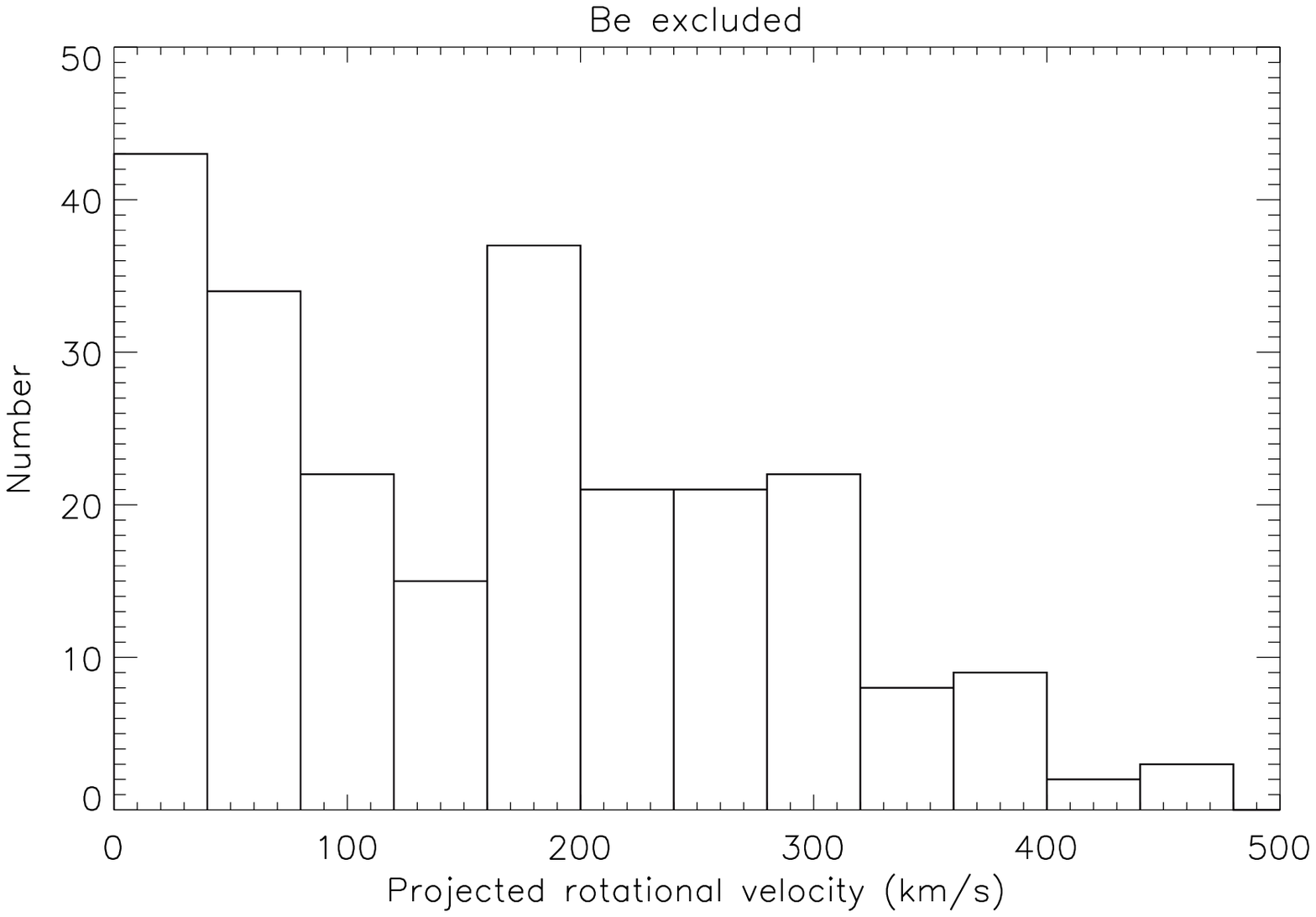}
   \caption{Histograms of the estimated projected rotational velocities, excluding stars of uncertain status. Top: all targets; Middle: excluding O9.5-type targets; Bottom: excluding Be-type candidates}
   \label{f_hist}
\end{figure}

In the bottom panel of Fig. \ref{f_hist}, the projected rotational velocity distribution is shown for the sample excluding the Be-type candidates. As might be expected, this leads to a decrease in the number of rapidly rotating stars, whilst the bimodality is still present. There remains a small number of rapidly rotating stars (\vsini$\ge$ 400 \kms), although we cannot exclude them having weak Be-type emission.

The median value of \vsini\ for the Be-star sample (excluding VFTS 102)  is 254~\kms\ (238~\kms\  excluding the stars of uncertain status), and the mean $249\pm11$~\kms\ ($244\pm12$~\kms). Although any bias introduced by the undetectability of weak stellar emission would presumably have skewed the results in favour of faster rotators, these values are somewhat smaller than the mean of $\sim$270~\kms\ for LMC field and cluster Be stars reported by \citet{mar06}.

\subsection{Comparison with previous studies} \label{d_comp}

\begin{figure}
   \centering
   \includegraphics[scale = 0.5,angle=0]{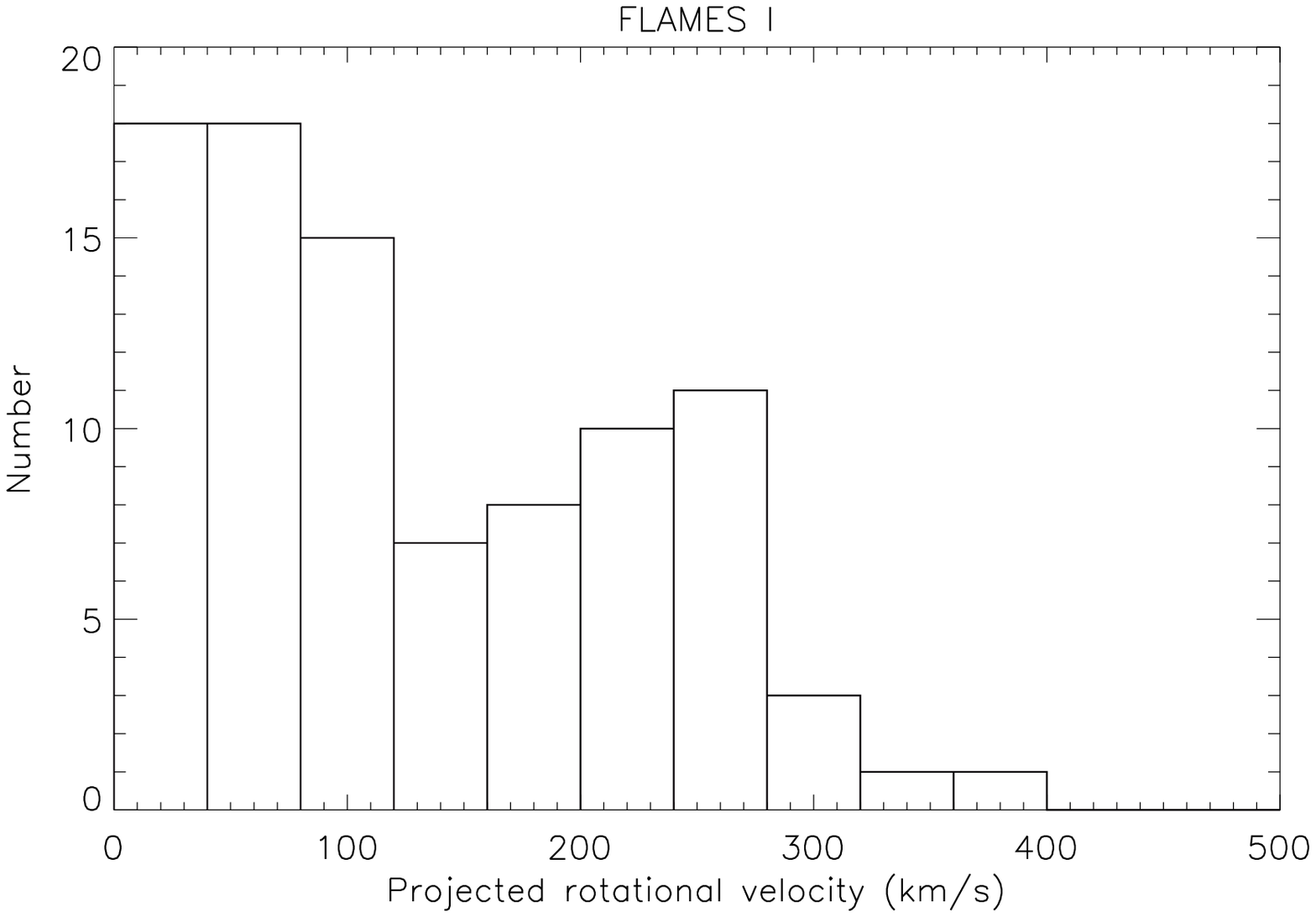}\\
   \includegraphics[scale = 0.5,angle=0]{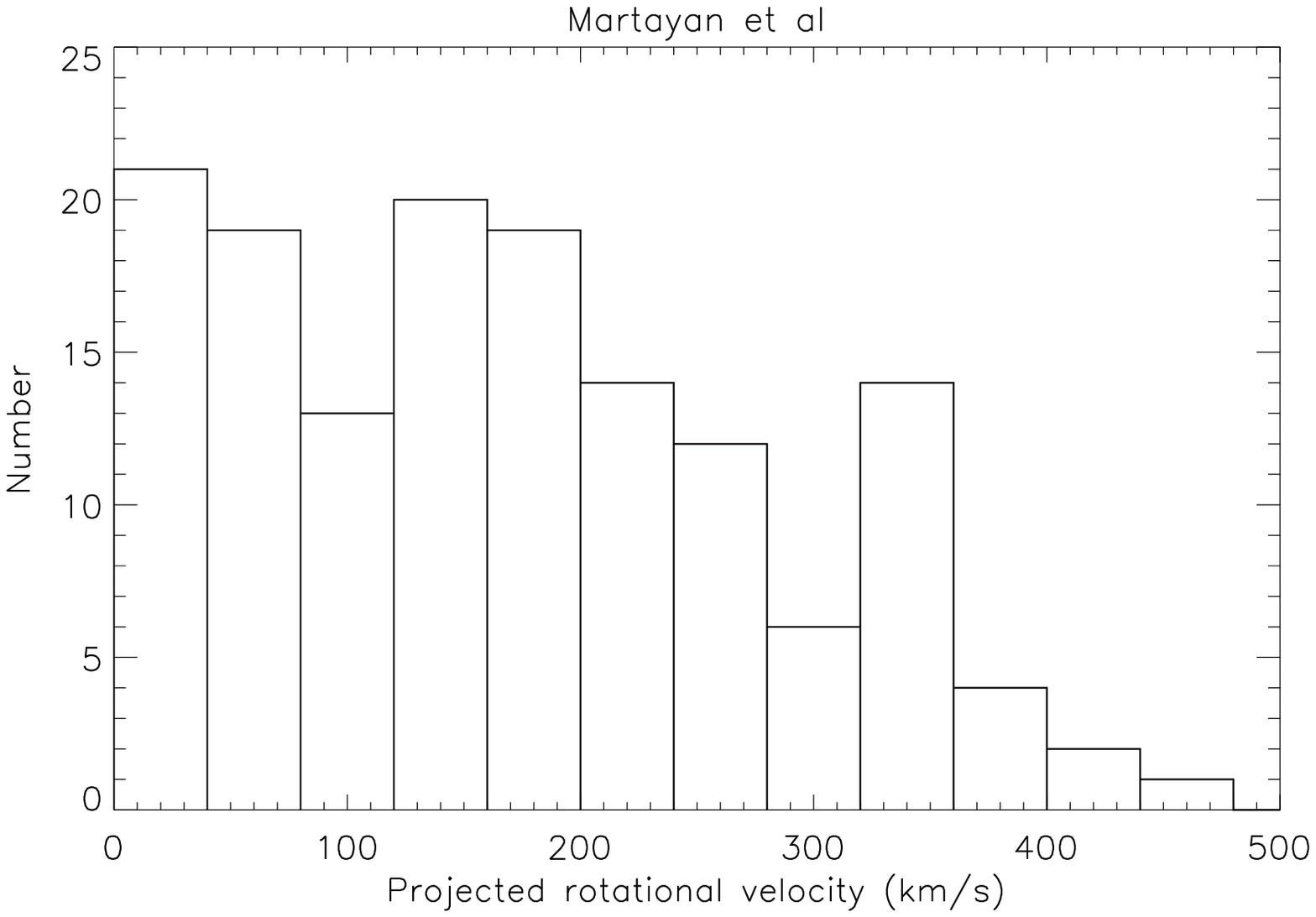}

   \caption{Histograms of the estimated projected rotational velocities for the surveys of \citet{hun08a} and \citet{mar06}.}
   \label{f_hist_comp}
\end{figure}

\begin{figure}
   \centering
   \includegraphics[scale = 0.5,angle=0]{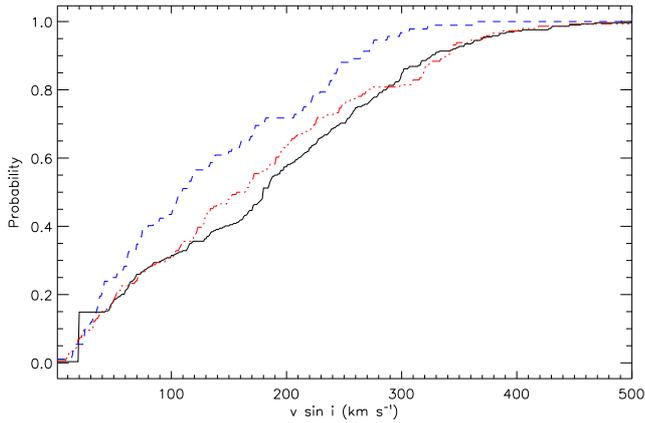}
   \caption{Cumulative distribution function for the our estimated projected rotational velocities (solid), together with those for the surveys of \citet{hun08a} (dashed) and \citet{mar06} (dot-dashed). For the current dataset, the unresolved spectra have been given an arbitrary \vsini\ of 20\, \kms}
   \label{f_cfp}
\end{figure}

Previous observational investigations of rotation in B-type stars have been discussed in the introduction. Here we limit the discussion to surveys in the Large Magellanic Cloud that are directly comparable with the current results.  The FLAMES-I survey observed targets towards two LMC clusters, N11 and NGC2004 and provided estimates of projected rotational velocities  \citep{hun08a}.  These estimates were not corrected for the systematic effect at near critical rotation discussed in Sect.  \ref{s_ft}. The rotational velocity distribution inferred by \citet{hun08a} had a modal value of 100\kms corresponding to $\Omega$/$\Omega_c\simeq 0.2$\ and hence we would expect that most of their sample will not contain significant systematic effects. In Figs. \ref{f_hist_comp}  and \ref{f_cfp}, the projected rotational velocity distribution and cumulative distribution function are shown for the 92 targets that had no evidence for binarity, were not supergiants and had a spectral type of O9.5-B3 (consistent with our current sample).
  
\citet{mar06} provided estimates for 106 B-type and 47 Be-type stars towards NGC2004. 
Supergiants and spectroscopic binaries were also excluded in this study, and most investigated stars had masses in the range 4\Msun to 20\Msun. Again limiting the sample to the 146 O9.5-B3 leads to the distributions shown in Fig. \ref{f_hist_comp} and  \ref{f_cfp}.

Neither of these two studies identified a bi-modal distribution. For example, \citet{hun08a} fitted a similar histogram (but with a different bin size and sample; their Fig.~9a) by a uni-modal distribution. However tailoring their sample to match our range of spectral types leads to evidence for a bi-modal distribution, although this must be considered as marginal given their small sample size. By contrast, the larger sample of \citet{mar06} shows no clear evidence for bi-modality, although this cannot be excluded.

To investigate whether these samples originated from the same parent population as our current sample, we have used the two sample Kolmogorov-Smirnov  and Kuiper tests. These implied that for
the sample of  \citet{hun08a},  this was unlikely (P $\sim 1\times 10^{-3}$\ and  $6\times 10^{-3}$\ respectively) but for that of  \citet{mar06}, the probability was relatively high (P $\sim 0.4$\ and 0.2 ). This is consistent with cumulative distribution functions plotted in Fig.~\ref{f_cfp}.

These comparisons must be treated with caution as the samples remain poorly matched. Firstly the selection criteria adopted for the three surveys will have introduced different observational biases. For example both the samples of  \citet{hun08a}  and \citet{mar06}  contained evolved main sequence stars (see their Figs.~7 and~5, respectively).  As the 30~Doradus region is thought to be young, our sample should contain a higher fraction of  relatively unevolved B-type stars, although we do not currently have spectral types or luminosity estimates to confirm this.  Secondly the previous studies observed a mixture of cluster and field stars, with the latter dominating. For example, \citet{mar06} estimated that $\sim$75\% of their sample were field stars; given the complex nature of the 30 Dor region, the fraction of field stars (in the traditional sense) in the current sample will be relatively low. Thirdly the sample of \citet{mar06} was obtained at two epochs and they only identified approximately 10\% of their sample as binaries.  By contrast for the sample of \citet{hun08a},  \citet{eva06} had identified 23\% of the NGC2004 targets and 36\% of the N11 targets as binaries and these have been excluded from the sample considered here. Hence given that the Tarantula survey has a lower limit of 37\% on the binary fraction \citep{dun11}, we conclude that the sample of \citet{mar06} probably contains a significant fraction of binaries. 


\section{Rotational velocity distribution}

The relatively large size of our sample has been exploited to investigate its intrinsic rotational velocity distribution. We first describe our methodology, and then outline our results and investigate their robustness.

\subsection{Deconvolution of the \vsini\  distribution} \label{d_decon}

If the stellar rotation axes are randomly distributed, it is possible to infer the probability density for the rotational velocity distribution (P(\ve)) from that for \vsini. Previously several authors have investigated this by assuming  an analytical function for P(\ve) and then optimising the fit with observation after convolution to allow for random inclinations. For example, for the analysis of the FLAMES-I datasets,  rectangular, Maxwellian \citep{mok06} and Gaussian \citep{duf06, mok06, hun08a} distributions were adopted. Given the structure in the projected rotational velocity distribution discussed above, it is not clear what adopted function would be appropriate. 

We have therefore attempted to deconvolve the projected rotational velocity distribution to directly infer that of the rotational velocities. We have adopted the iterative procedure of \citet{luc74} which if it converges will converge to the point of maximum likelihood. For the continuous function, representing the observed projected rotational velocities ($\tilde{\phi}$\ in the notation of Lucy), we followed his recommended procedure. The initial estimate for the rotational velocity distribution ($\psi^{0}$\ in the same notation) was taken to be a Gaussian with a positive abscissa; tests showed that this choice did not affect the final iterative solution, consistent with the discussion of \citet{luc74}. 

The effectiveness of this procedure has been tested by taking simulated distributions for P(\ve) and then using a Monte-Carlo technique to generate P(\vsini). These were then de-convoluted to obtain an estimate for P(\ve). Simulated distributions considered included Gaussian, triangular and step functions. In general the deconvolution retrieved the large scale structure, while the estimated {\em projected} rotational velocity distribution, ($\tilde{\phi}$ in Lucy's notation), agreed well with the simulated P(\vsini). Additional small scale structure was present especially where the adopted rotational velocity distribution was discontinuous. Also structure varied from simulation to simulation and appeared to be connected with that due to the Poisson noise. The only significant concern to arise in these simulations was that on occasions a small excess of slowly rotating stars were found even though there was excellent agreement between the simulated projected rotational velocity distribution, P(\vsini), and that inferred from the deconvolution procedure ($\tilde{\phi}$). In turn this implied that relatively small errors in the observed projected rotational velocity distribution could propagate through to that for the rotational velocities, P(\ve).

Our adopted procedure is similar to that used by \citet[and references therein]{zor12} to investigate the rotational velocity distributions for Galactic A-type stars. However we differ from \cite{zor12} in not preprocessing our observational projected rotational distributions using a Gaussian filter. Our approach will not remove spurious structure in our inferred
rotational velocity distribution due to the Poisson noise in the observational distributions. However it will also illustrate the limitations of this approach and its sensitivity to the assumptions made.

\begin{figure}
   \centering
   \includegraphics[scale = 0.5,angle=0]{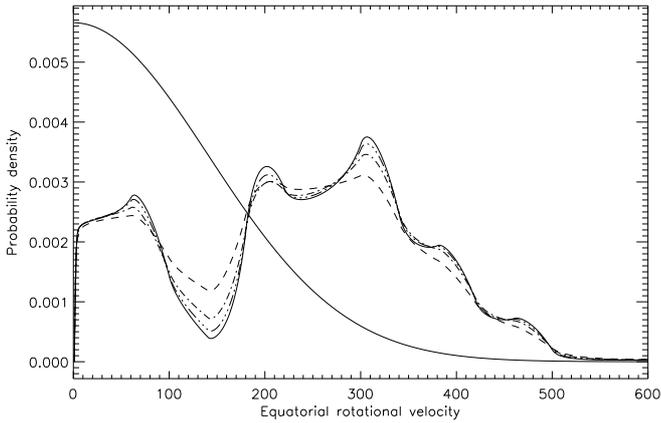}

   \caption{Distribution of de-projected rotational velocities from our iterative procedure for the observed \vsini\ distribution binned to 40 \kms. The initially assumed Gaussian distribution (solid line) and the results after two (dashed), four (dot-dash), six (double dot-dashed) and eight (solid) iterations are shown for \ve$\leq 500$\ kms.} 
   \label{f_v_40}
\end{figure}

The choices of the number of iterations and the bin size for the observed projected rotational velocity are important. For the former, the iteration initially converges rapidly and then tends to accentuate small scale structure that may not be real. \citet{luc74} in his two rotational velocity examples only considered two iterations. In Fig. \ref{f_v_40}, we show (for an observation bin size of 40 \kms), the originally assumed distribution of the rotational velocities and the solutions after two, four, six and eight iterations. The solution after two iterations shows the major features of the  distribution with further iterations introducing additional structure  at approximately at 70, 270 and 420\kms. Although the signal will not be band-limited, this structure would appear to contain frequencies higher than the Nyquist frequency (or equivalently in the rotational velocity domain smaller than half the bin size). Additionally the errors due to Poisson statistics map onto a typical uncertainty of approximately 5$\times 10^{-4}$ in P(\ve) and hence this additional structure is probably not significant. 

\begin{figure}
   \centering
   \includegraphics[scale = 0.5,angle=0]{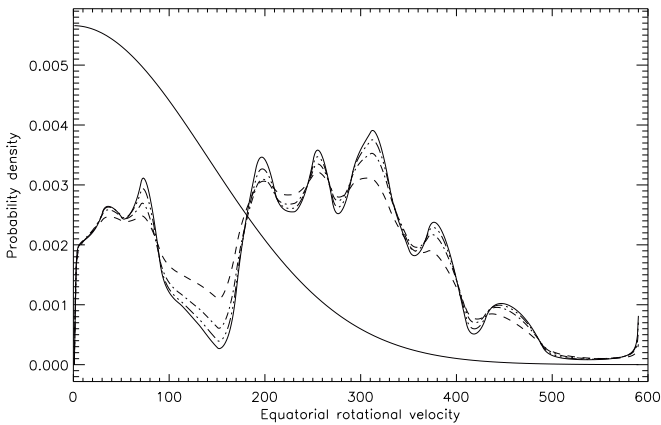}

   \caption{Distribution of de-projected rotational velocities from our iterative procedure for the observed \vsini\ distribution binned to 20 \kms. Notation is identical to Fig. \ref{f_v_40}.}
   \label{f_v_20}
\end{figure}

In the case of the adopted observational bin size it is necessary to balance the resolution in the velocity domain against the (Poisson) noise in the star counts. We have taken a pragmatic approach by considering two different possible bin sizes  of 20 and 40 \kms, with for the former the spectrally unresolved targets being equally distributed between the first two bins. The results are illustrated in Figs. \ref{f_v_40} and  \ref{f_v_20} for different number of iterations. The large scale structure of the estimates of the rotational velocity attribution are similar but the smaller bin size leads to more small scale structure. Although this will to some extent reflect the higher sampling rate, we believe that mainly that it arises from random errors in the \vsini\ statistics. 

 On the basis of the above tests, we have adopted the larger bin size together with 4 iterations in our analysis and discussion given below. We note that the major conclusions would not have been significantly affected by other choices.

\subsection{Results}

\begin{figure}
   \centering
   \includegraphics[scale = 0.5,angle=0]{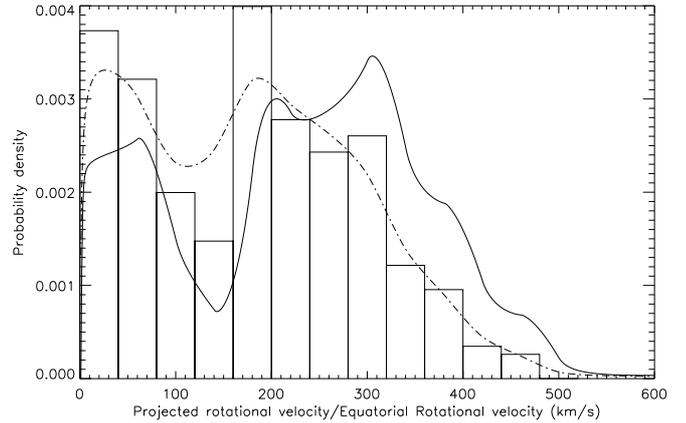}

   \caption{Histogram of observed normalised \vsini\ distribution binned to 40 \kms. Also shown are the estimates after 4 iterations of the probability densities for the projected rotational velocity distribution, P(\vsini) ($\tilde{\phi^{4}}$ in Lucy's notation -- dot-dashed line) and for the rotational velocity, P(\ve) ($\psi^{4}$\ in Lucy's notation-- solid line).}
   \label{f_vdist}
\end{figure}

In order to illustrate our results, Fig. \ref{f_vdist} shows the observed \vsini\ histogram from Fig. \ref{f_hist}, normalised to  unity when integrated over velocity. Also shown are the estimates after four iterations of the probability density of the projected rotational velocity distribution ($\tilde{\phi^{4}}$\ in Lucy's notation) and of the corresponding probability density for the rotational velocity, P(\ve) ($\psi^{4}$\ in Lucy's notation). There is good agreement between the observed \vsini\ distribution and that inferred ($\tilde{\phi^{4}}$) from the iterative process. Additionally the estimate of the rotational velocity distribution appears reasonable with the large scale structure of the \vsini\ distributions being conserved but moved to larger velocities due to the effect of inclination.

In Table \ref{t_veq} our best estimates of the probability distribution function are tabulated. For smaller rotational velocities, the estimates  have  been rounded to 5$\times 10^{-5}$\ but for larger velocities a small interval was chosen. We emphasis that these values are subjective as they require a judgement about the validity of the smaller scale structure. Also listed in this table is the cumulative distribution function (cdf)  corresponding to this best estimate and because this is truncated at \ve$\leq 600$\,\kms, it does not reach unity. We recommend that:
\begin{enumerate}
\item either the range up to the critical velocity is populated
\item or that the cdf and estimate of P(\ve) are renormalised
\end{enumerate}
depending on the nature of the application.

\begin{table}
\caption{Estimates of the probability density of the rotational velocity, P(\ve), and its cumulative distribution function, cdf.}
\label{t_veq}
\begin{center}
\begin{tabular}{rccccll}
\hline\hline
\ve   & P(\ve)$\times 10^3$  & cdf   \\
\hline
0      &      2.20   & 0        \\   
20    &      2.40   & 0.046 \\
40    &      2.45   & 0.094\\
60    &      2.55   & 0.144 \\
80    &      2.25   & 0.192 \\
100  &      1.40   & 0.239 \\
120  &      1.00   & 0.253 \\
140  &      0.75   & 0.270 \\
160  &      1.05   & 0.288 \\
180  &      2.40   & 0.322 \\
200  &      3.00   & 0.377 \\
220  &      2.85   & 0.435 \\
240  &      2.75   & 0.492 \\
260  &      2.90   & 0.548 \\
280  &      3.20   & 0.609 \\
300  &      3.30   & 0.674 \\
320  &      3.20   & 0.739 \\
340  &      2.50   & 0.796 \\
360  &      2.00   & 0.841 \\
380  &      1.80   & 0.879 \\
400  &      1.50   & 0.912 \\
420  &      1.10   & 0.938 \\
440  &      0.75   & 0.956 \\
460  &      0.70   & 0.971 \\
480  &      0.50   & 0.983 \\
500  &      0.21   & 0.990 \\
520  &      0.09   & 0.993 \\
540  &      0.08   & 0.995 \\
560  &      0.06   & 0.996 \\
580  &      0.04   & 0.997 \\
\hline
\end{tabular}
\end{center}
\end{table}

All the de-convolutions show a double peak in the  equatorial velocity distribution consistent with that observed in the projected rotational velocity distribution.  Approximately one quarter of the sample have a rotational velocity of less than 100 \kms\ and there appears to be another peak 250-300 \kms\ although this is complicated by the varying degree of small scale structure found in the different de-convolutions. 

We have attempted to fit the de-convolved distribution using analytical functions. Initially two Gaussian profiles were adopted but these were found to give a relatively poor fit to our estimated distribution.  Two Maxwellian functions as discussed by \citet{zor12} were also considered but would not have reproduced the significant value of P(\ve) as \ve$\rightarrow 0$. We therefore recommend that either the values listed in Table \ref{t_veq} or  analytical fits appropriate to the specific applications are adopted.

The most distinctive feature of the \vsini\ distribution is its bi-modal nature, which is also present in the estimation of the rotational velocity distribution, illustrated in Fig. \ref{f_vdist}. Although our sample contains approximately 300 targets, it will still be subject to significant random sampling errors due to its finite size. In Fig. \ref{f_hist}, the population of the two 40 \kms\ bins with the lowest projected rotational velocities have values of approximately 40 corresponding to an estimated standard error of approximately 6. Hence even decreasing the populations of both these bins by twice this estimated error (note that if these bins had been overpopulated our estimate of the standard error would also be too high) would not remove this bi-modal behaviour. Additionally Kolmogorov-Smirnov tests using uni-modal distributions (for example a single Gaussian fit to our estimated rotational velocity distribution) lead to very small probabilities that they were the parent populations. Hence we conclude that the projected rotational velocity distribution (and hence the underlying rotational velocity distribution) is bi-modal, although we accept that our sample size limits the information on the structure of the two components.

In Sect. \ref{d_comp}, we discussed the projected rotational velocities in two other LMC samples. Those of \citet{hun08a} showed some evidence for a bimodal distribution but that of \citet{mar06} appeared unimodal. However given the differences in the samples in terms of age, fraction of field stars and undetected binaries, we do not consider this discrepancy to be significant.

\section{Origin of bi-modal distribution} \label{two_peaks}

The bi-modal distribution of rotational velocities implies that our stellar sample is composed of, 
at least, two different components. Which physical processes lead to the existence of these 
two components? Why do most B-type main sequence stars rotate fast, while about a quarter or so 
rotate slowly? Whatever the answers to these questions, they are not contained in the standard 
evolutionary theories of single stars, which predicts
only small changes of the surface rotational velocity during core hydrogen 
burning \citep[see, for example,][]{bro11a}.

The two components, which we find, could either correspond to different ages, different star formation conditions, or emerge as a consequence of differences in the evolution of stars (e.g., in close binaries). We emphasize again that age differences can only lead to the difference in the mean rotation rate of the two components unless the standard stellar evolution picture is wrong or incomplete. Differences in age and star formation conditions could produce stellar components with different spatial or kinematic properties. We inspect our data in this respect in the next two subsections. Thereafter, we discuss the possible impact of non-standard stellar evolution.

\subsection{Spatial variations} \label{d_spat}

\begin{figure}
   \centering
   \includegraphics[scale = 0.5,angle=0]{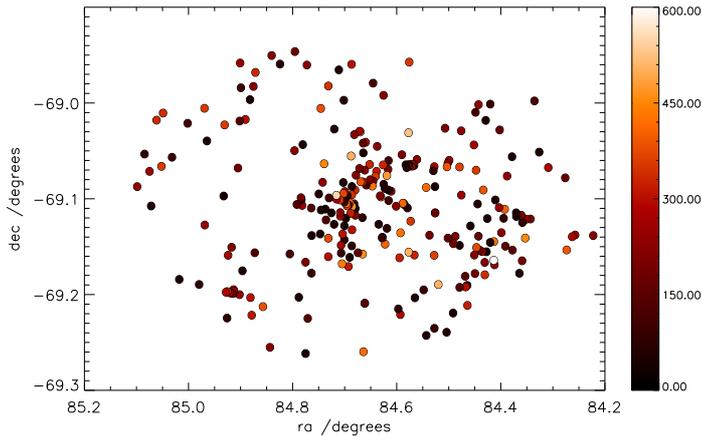}

   \caption{Spatial position of our targets with different colours used to identify different ranges of projected rotational velocity, according to the colour bar at the right side. The relatively old cluster Hodge 301 lies at a right ascension of 84.50$^\circ$ and a declination of -69.06$^\circ$.}
   \label{f_spatial}
\end{figure}

The 30~Doradus region has had a complex star-formation history over the past few tens of Myrs, as demonstrated by the five distinct stellar groups identified by \cite{wal97}.  For instance, the VFTS 
observations include targets in the cluster Hodge~301, which is $\sim$20\,Myr older than the central parts of 30~Doradus \citep{gre00}. Indeed, even in the central 30~Doradus cluster (NGC\,2070), \citet{sel99}  identified three distinct temporal components, with \citet{sab12} recently identifying two spatially-distinct components which appear to have an age difference of $\sim$1\,Myr.
An obvious test is thus to check whether the slow rotators concentrate in particular regions of the  VFTS field. In Fig \ref {f_spatial}, the spatial positions of the sample of our targets are illustrated,
together with their estimated projected rotational velocity. The concentrations of stars at
specific positions follows directly from the morphology of the region.

From this figure, there appears to be no evidence for any spatial segregation with projected rotational velocity. For example,  the percentage of targets with values of \vsini\ less than 100\,\kms\ within 0.1 degrees of the central concentration of stars associated with Hodge 301 is 33\%, in excellent agreement with the 33\% found for the whole sample. 


To further assess any spatial variation, the sample was split into five velocity bins, each of 100\kms, except the top bin which encompassed all objects with \vsini $\geq 400$\kms. Cumulative radial-distribution functions were calculated for each bin. Two-sample Kolmogorov-Smirnov (KS) tests were then performed on all possible pairs of the radial-distribution functions. No statistically significant variation was found between the pairings, suggesting the spatial distribution does not influence the \vsini\ distribution.

In this context it is interesting to consider the results from \citet{wol08} for projected rotational velocities of 24 OB-type stars in R136. Their sample stars were located within $\sim$25$''$ of the centre of R136 (equivalent to a projected distance of $\sim$6\,pc), i.e. the majority of their stars lie within the dense central part of the FLAMES field that was not observed with the Medusa fibres. From comparisons with results for other samples in the LMC (including those from Hunter et al. 2008b), Wolff et al. concluded that the mean rotational velocity of the massive stars in R136 is larger than that found from other samples in the LMC, both for stars in (less dense) clusters and in the field population.

\citet{wol08} noted that R136 lacks the population of very slowly-rotating stars seen elsewhere (and which also features in the results presented here). They divided their sample into high- and low-mass bins, with inferred evolutionary masses in the range of 15\,$\le$\,$M_\ast$\,$\le$\,30\,$M_\odot$ and 6\,$\le$\,$M_\ast$\,$\le$\,12\,$M_\odot$, respectively.  The majority of the stars in the lower-mass bin lie below the faint magnitude cut-off for the VFTS, but the lack of slow rotators is also evident in their high-mass sub-sample (see their Fig.~11) -- they argued that these results might reflect the initial conditions within which the stars formed, i.e. the relationship of rotational velocity with stellar density.  As noted above, we see no compelling spatial trends in the results from our significantly larger VFTS sample, with the caveat that our results do not include stars in the central few pc of R136.

\subsection{Radial velocity variations} 

A bi-modality found in Galactic targets by \cite{hua10} has been interpreted as due to a field star and a cluster population. Although we see no evidence for this in the spatial distribution of targets, we have also investigated the estimated stellar radial velocities of our sample. These have been  measured Kennedy et al. (in prep.) using profile fitting to selected spectral features with a methodology similar to that discussed by \citet{san12b}. We have divided our sample into two groups with projected rotational velocities less than or greater than 100 \kms. The former will be a mixture of slowly rotating stars and those viewed at low angles of inclination; however the contamination should be relatively small given that the mean value of $\sin i$\ for random inclinations is approximately 0.78. The latter will be stars with rotational velocities in excess of 100 \kms.

\begin{table}
\caption{Average radial velocities for sub-samples with $vsini$ smaller and larger
than 100\,km/s (cf. Fig~\ref{f_hist}), compared to the full sample.}
\label{t_vr}
\begin{center}
\begin{tabular}{lcc}
\hline\hline
\vspace*{-0.2cm}\\
Sample                    &   Number & $\langle v_{\rm rad} \rangle$\ \kms  \\
\hline
\vspace*{-0.2cm}\\
\vsini $< 100$\,\kms      &   89     &  269$\pm$17  \\
\vsini $\geq100$\,\kms    &   172    &  267$\pm$18  \\
\\[-4pt]
All                       &   261    &  267$\pm$17 \\
\hline
\end{tabular}
\end{center}

\end{table}

In Table \ref{t_vr}, we summarize the mean and standard deviations for the different samples, together with the sample sizes. Although reliable values were not available for approximately 10\% of the targets, we do not expect this to introduce a significant bias. For the two samples, both the means and standard deviations are effectively identical. Coupled with the results presented in Sect. \ref{d_spat}, we conclude that there is no evidence that the bi-modality arises from different populations, due to either sequential star formation or from contamination with field stars.

\subsection{Binary evolution}

The fraction of Galactic massive close binaries is high, i.e.,  at least 50\% \citep{san12a}. Similar high fractions have been found for the O- and B-type VFTS samples \citet[][cf., Sect.~\ref{s_obs}]{san12b}. Since the time evolution of the rotation of massive main sequence stars can be strongly affected by the interaction with a companion \citep{lan12}, this could in principle affect the distribution of rotation rates of our sample. The reason is that, while radial velocity variables have been eliminated from our sample, \citet{deM11,deML12} found that the vast majority of post-interaction objects are either single stars or appear as such.

De Mink et al. (2012b) showed that for constant star formation, the fraction of apparently single post-binary interaction objects can be as high as 30\%. This fraction would be considerably larger in a sample in which radial velocity variables are removed, as except for a small fraction of Algol-type systems, those are almost exclusively pre-interaction binaries. For the current sample, for which the age distribution is not yet known, on one hand the fraction of apparently single close binary products is expected to be smaller than 30\% due to the expected presence of early B-type stars which are coeval with the core region of 30 Doradus. On the other hand, this fraction is enhanced due to the removal of the radial velocity variables from our sample.  In conclusion, at present a substantial contribution of stars that are the product of binary evolution, can not be excluded.

The binary population synthesis models of  \citet{deMP12} imply that  binary interactions, i.e., mass and angular momentum transfer and stellar mergers, produce a population of fast rotators, with rotational velocities up to about 600 km/s.   In principle, binary interaction can also produce slow rotators, for example, as a result of spin down by tides. However, \citet{deMP12} find that the effect of this on a stellar population is very small.   The low velocity peak of our distribution (see Fig. 8) contains about 25\%, which is too many to be explained solely as the products of binary evolution.  Indeed, to produce the low velocity component of our distribution by close binary effects could require the hypothesis that mass transfer and/or mergers lead to significant magnetic fields in the binary remnant (cf., Sect. 5.4) as supported by the recent discovery by \citet{gru12a} of a strong magnetic field in the companion of Plaskett's star.

\subsection{Magnetic fields} \label{s_B}

\citet[and references therein]{zor12} have undertaken an extensive study of the rotational velocities in Galactic late B-type and A-type stars, using a similar methodology to estimate rotational velocity distributions. Their large dataset allowed them to study variations as a function of age and mass. They find that stars with masses $\la 2.5$\,\Msun\ have a uni-modal distribution of rotational velocities, whilst the more massive stars (2.5-3.9\Msun) show a bi-modal structure. \citet{zor12} fit the latter with two Maxwellian functions and find that the percentage of targets in the low velocity Maxwellian varies from 8-20\% depending on the mass range considered. Additionally they find that the peak of the higher velocity Maxwellian lies in the range 200$\la$\ve$\la$240\,\kms\ for the highest masses they consider.

These results are not directly comparable with those presented here since \citet{zor12} 
sampled different field and cluster environments. Additionally our sample is not large enough 
to be subdivided into smaller mass ranges, while we do not have age estimates for our targets. However, both the fraction of slow rotators (25\% for our fit; $8-20$\% in the higher mass samples of \citeauthor{zor12}) and the peak value of the high velocity component  
(280 \kms\  and 200-240 \kms\, respectively) are similar. 

\cite{zor12} discuss two possible explanations for the bi-modality in their distributions. The first
is that a fraction of stars loses angular momentum via tidal breaking within a binary system.
For our sample, we can exclude this scenario, as \citet{deML12} show that the number
of affected stars is much too small. The second possibility discussed by \cite{zor12} 
is that all stars were born rotating rapidly, but that a fraction of them had or acquired magnetic fields. Indeed, the fraction of magnetic main sequence stars in the mass range 2.5-3.9\Msun is believed  to be about 15\% \citep{don09}, whilst their rotation may be slowed down by magnetic breaking.

It has become evident in recent years that magnetic fields could potentially play a significant role in the evolution of some massive stars. In the LMC sample of early B-type stars discussed in Sect.~\ref{d_comp}, \citet{hun08b} found a sub-sample of 15\% of their stars with slow rotation and high nitrogen enrichment. While direct magnetic field measurements in these stars are not yet feasible, \citet{mor08} and \citet{hub11} have identified Galactic analogues for which the incidence of magnetic fields is indeed high. \citet{gru12b}  find magnetic fields in $\sim$10\% of their early B~stars.

\citet{mey11} and \citet{pot12b, pot12a} have explored stellar evolution models for magnetic massive main sequence stars. Based on a model for magnetic braking of \cite{udd02}, they both find that magnetic massive main sequence models can in fact spin-down for reasonable field strength. \citet{pot12b}  performed  a population synthesis matched to the B-type stellar sample of \citet{hun08b}. They produced a bi-modal velocity distribution, with a significant component of slow rotators ($v_e < 60$\,km/s). However their distribution of fast rotators peaks at $v_e \simeq 100$\,km/s (possibly due to their choice of the initial rotational velocity distribution), compared to $v_e \simeq 250$\,km/s for our sample (Fig.~\ref{f_vdist}). 

Of course, if magnetic spin down was responsible for the low-velocity component in our
rotational velocity distribution, one may wonder about the origin of the magnetic fields in these main sequence stars. Three possibilities are being discussed in the literature. The fields could 
be fossil, i.e. inherited from the star formation process \citep{don09}. Alternatively, the fields could be generated by strong binary interaction, i.e. binary merger, mass transfer or common evelope evolution \citep{fer09, lan12}.  Finally, the fields could be generated by a dynamo process inside the the main sequence star \citep{spr02,pot12a}. We emphasize that these three alternatives predict different time-dependances of the stellar rotation and magnetic field incidence, which might be testable if the age distribution of our sample stars can be established.

An additional important  braking mechanism could be mass loss via stellar winds. Winds are thought to be ubiquitous amongst massive stars, and despite the spectroscopic evidence for high-speed outflows being most prevalant amongst the most luminous O-type objects (in the form of P Cygni ultraviolet lines), they are also observed in the lower mass B-type stars. Although mass-loss rates are higher in O stars, wind braking \citep{lan98} could be relevant  for B-type stars given their longer evolutionary timescales. One interesting possibility is that of bi-stability braking \citep{vin10}, which might lead to a population of both slow and fast rotators, and hence play a role in the observed bi-modality. This could be tested when the stellar parameters of our sample stars have been established.
 
\section{Conclusions}

From the VFTS B-type stellar sample (approximately 540 stars),  we have constructed a sample 289 early B-type stars (plus 45 stars of uncertain binary status) by removing radial velocity variable stars and supergiants. These will have masses in the range 6 to 16 \Msun and are centered on the 30 Doradus region in the LMC. We have estimated their projected rotational velocities, and find that their distribution shows a distinct bi-modality, which neither correlates with position in the field nor with radial velocity. 

We have deconvolved the \vsini\ distribution to obtain the most likely underlying distribution
of intrinsic rotational velocities. This shows a low velocity component containing
about one quarter of the sample stars, with $v_{\rm e} <$\,100\,km/s, and a second component of stars rotating predominantly with velocities in the range 200\,km/s$ < v_{\rm e} <$\,350\,km/s.

This distribution is not inconsistent with previous LMC surveys, despite a bi-modality having not been clearly demonstrated before. A conclusive interpretation of our finding is not yet possible, in particular because the age distribution and chemical composition of the sample stars are still unknown. An origin of the bi-modality due to evolutionary effects relating to binarity and/or magnetic fields or that the velocity distribution was inherited in the star formation process remain possibilities.

\begin{acknowledgements}
We are grateful to Paul Crowther, Morgan Fraser, Ian Hunter and Danny Lennon for advice and comments on this paper. PRD thanks the Northern Ireland, Department of education and Learning for financial support. IB acknowledges support by the Austrian Science Fund (FWF). SdM is supported by NASA through Hubble Fellowship grant HST-HF-51270.01-A awarded by the STScI, operated by AURA, Inc., for NASA, under contract NAS 5-26555. SS-D acknowledge financial support by the Spanish Ministry of Economy and Competitiveness (MINECO) under grant number AYA2010-21697-C05-04  and the Consolider-Ingenio 2010 Program grant CSD2006-00070 (http://www.iac.es/consolider-ingenio-gtc). 
\end{acknowledgements}
 
\bibliography{./literature.bib}
\end{document}